\documentclass[
reprint,
amsmath,amssymb,
aps, prl,
]{revtex4-2}

\usepackage[title]{appendix}
\usepackage{graphicx}% Include figure files
\usepackage{dcolumn}% Align table columns on decimal point
\usepackage{bm}% bold math
\usepackage{bbold}
\usepackage[hidelinks]{hyperref}
\usepackage[capitalise]{cleveref}
\usepackage{lipsum}
\usepackage[dvipsnames]{xcolor}
\usepackage{amsfonts}
\usepackage{yfonts}
\usepackage{booktabs}

\usepackage{amsthm}
\usepackage[english]{babel}
\usepackage[scr=boondoxo,scrscaled=1.05]{mathalfa}
\theoremstyle{definition}

\newcommand{\rmd}{\mathrm{d}}

\newcommand{\PCell}[2]{\parbox[t]{#1}{\raggedright #2}}

\DeclareFontFamily{U}{futm}{}
\DeclareFontShape{U}{futm}{m}{n}{
  <-> s * [.97534] fourier-bb % but changing the magnification factor
  }{}
\DeclareMathAlphabet{\mathbbs}{U}{futm}{m}{n}

\begin{document}

\preprint{APS/123-QED}

\title{Mutual Linearity in Nonequilibrium Langevin Dynamics}
\author{Jiming Zheng} \email{jiming@unc.edu}
\author{Zhiyue Lu} \email{zhiyuelu@unc.edu}
\affiliation{Department of Chemistry, University of North Carolina-Chapel Hill, NC}

\date{\today}

\begin{abstract}
Understanding how nonequilibrium systems respond to perturbations is a central challenge in physics. In this work, we establish mutual linearity in nonequilibrium overdamped Langevin systems. This theory provides a framework for controlling and designing nonequilibrium responses in continuous systems. When a dynamical parameter is locally perturbed at a single position, the stationary densities at any two positions are linearly related. It further leads to mutual linearity among different stationary state-current observables. We also extend the mutual linearity to non-stationary relaxation processes in the Laplace domain. Our theory reveals that mutual linearity in both discrete and continuous systems originates from the same one-dimensional response structure. We further show that mutual linearity is robust under finite-width perturbations. As an application, we demonstrate the mutual linearity and its finite-width robustness in the F$_1$-ATPase rotary motor model. 
\end{abstract}

\maketitle

\emph{Introduction.}--- Understanding how nonequilibrium systems respond to external perturbations is a central challenge in statistical physics, biophysics, and chemical physics. Near equilibrium, the fluctuation-dissipation theorem \cite{kubo1966fluctuation} links linear response to spontaneous fluctuations. Far from equilibrium, no comparable universal structure exists in general. Stochastic thermodynamics \cite{seifert2012stochastic,peliti2021stochastic} has provided a trajectory-level framework for nonequilibrium dynamics and dissipation, motivating a broad class of fluctuation-response relations \cite{maes2020response,maes2013fluctuation,baiesi2009fluctuations,bao2024nonlinear,aslyamov2025nonequilibrium,ptaszynski2026nonequilibrium,chun2026fluctuation,aslyamov2026dynamical}, response inequalities \cite{aslyamov2026dynamical,aslyamov2025nonequilibrium,ptaszynski2026nonequilibrium,chun2026fluctuation,zheng2025nonlinear,zheng2025unified,zheng2025universal,zheng2026thermodynamic,kwon2025fluctuation,dechant2025finite,dechant2020fluctuation,hasegawa2019uncertainty,gu2026spectral}, and macroscopic response theories \cite{aslyamov2026macroscopic,zheng2025nonequilibrium}. These response theories now play an important role in active matter \cite{dal2019linear,davis2024active}, chemical reaction networks \cite{ueon2026transiently,chun2023trade}, and biophysical systems \cite{tu2008modeling,mattingly2026mechanical}.

Among nonequilibrium response relations, mutual linearity in Markov networks is particularly striking \cite{harunari2024mutual,floyd2025local,bebon2026mutual,zheng2026mutual}: when a single transition rate is perturbed, different observables can be affinely related even though each observable may nonlinearly depend on the perturbed rate. This structure has been interpreted through first-passage times \cite{khodabandehlou2025affine}, extended to multi-edge perturbations \cite{dal2025mutual}, and recently connected to trajectory-level response theory \cite{zheng2026mutual}. However, the above existing theories suffer from a significant drawback from the application perspective: the mutual linearity is formulated for discrete systems. By contrast, experiments are frequently performed on continuous systems, including active matter \cite{mizuno2007nonequilibrium,turlier2016equilibrium,mizuno2008active}, molecular motors \cite{toyabe2010nonequilibrium,gieseler2021optical}, or levitated nano-particles \cite{gonzalez2021levitodynamics,hempston2017force}. It is not straightforward to extend existing theories to such continuous systems.

In this letter, we show that mutual linearity is not restricted to discrete Markov networks, but also emerges in nonequilibrium overdamped Langevin dynamics. The key mechanism is a one-dimensional response geometry: a local perturbation excites a single response mode, and this mode remains parallel to itself as the perturbation strength varies. As a result, stationary densities at different positions obey affine relations, and the same structure extends to excluded-point state-current observables and to non-stationary relaxation dynamics in the Laplace domain. This continuous theory also reveals the unified origin of mutual linearity in both discrete and continuous Markov systems. Finally, we show that the affine structure is robust under finite-width local perturbations and demonstrate the theory in an F$_1$-ATPase model, providing a realistic route toward experimental tests of mutual linearity in continuous nonequilibrium systems.

\emph{Setup.}--- For illustrative purpose, let us consider one-dimensional overdamped Langevin systems. The minimal model captures the mutual linearity and its underlying response structure. Extensions to multi-dimensional cases are straightforward and discussed in the Supplemental Material \cite{supp}. We describe the one-dimensional overdamped Langevin system by
\begin{equation}
  \dot{x}_t=\mu(x_t)F(x_t)+\sqrt{2\mu(x_t)T(x_t)} \,\star\, \xi_t ,
  \label{eq:langevin}
\end{equation}
where $x_t$ is the stochastic state variable, $\mu(x)$ is the position-dependent mobility, $F(x)$ is the systematic force, and $T(x)$ is the local temperature field. The symbol $\star$ indicates the anti-It\^o convention required for thermodynamic consistency \cite{lau2007state}. The noise $\xi_t$ is a Gaussian white noise with a zero mean and an autocorrelation function $\langle \xi_t\xi_{t'}\rangle=\delta(t-t')$. Except otherwise specified, we set the Boltzmann constant to unity.

The probability density $p(x,t)$ associated with \cref{eq:langevin} obeys the Fokker--Planck equation
\begin{equation}
  \partial_t p(x,t) = -\partial_x \! \left[ \mu(x)F(x)p(x,t)-\mu(x)T(x)\partial_x p(x,t) \right].
  \label{eq:fp}
\end{equation}
By adopting the probability-current operator $\mathcal{J} = \mu(x) \bigl[F(x)-T(x)\partial_x\bigr]$ and the Fokker--Planck generator $\mathcal{L} = -\partial_x \mathcal{J}$,
\cref{eq:fp} can be compactly written as $\partial_t p(x,t)=\mathcal{L}p(x,t)$. Assume that the dynamics admits a unique stationary density $\pi(x)$ satisfying $\mathcal{L}\pi(x)=0$, and denote the corresponding stationary current by $j_{\mathrm{ss}}(x)=\mathcal{J}\pi(x)$.

We focus on state-current observables of the form
\begin{equation}
  Q(t) = \int \rmd x\,a(x)p(x,t) + \int \rmd x\,b(x)j(x,t),
  \label{eq:observable}
\end{equation}
where $a(x)$ and $b(x)$ are arbitrary smooth functions. This class of state-current observables encompasses a broad range of experimentally relevant quantities. Examples include state occupations $(a \neq 0,~ b = 0)$, local currents $(a = 0,~ b \neq 0)$, input power $(a = 0,~ b = F_{\mathrm{drive}})$, and entropy production $(a = 0,~ b = F_{\mathrm{tot}}/T)$. In the steady state, $p(x,t) = \pi(x)$ and $j(x,t) = j_{\mathrm{ss}}(x)$, so \cref{eq:observable} reduces to
\begin{equation}
  Q_{\mathrm{ss}} = \int \rmd x\,a(x)\pi(x)+\int \rmd x\,b(x)j_{\mathrm{ss}}(x).
  \label{eq:observable_ss}
\end{equation}

\emph{Elementary mutual linearity.}--- We fix one of the dynamical parameters $\phi(x) \in \{ F(x), \mu(x), T(x) \}$, and consider the perturbation localized at $z_0$ under a single parameter $\lambda$:
\begin{equation}
  \phi_{\lambda}(x) = \phi_{0}(x) + \lambda \, \delta(x - z_0).
\end{equation}
The delta function restricts the perturbation at a single point, and can be realized as the narrow limit of a Gaussian packet. For mobility $\mu(x)$ and temperature $T(x)$ perturbations, $\lambda$ is restricted so that the regularized $\mu_{\lambda}(x)$ and $T_{\lambda}(x)$ remain positive.

Let $\pi_{\lambda}(x)$ denote the stationary density of the perturbed dynamics. Our first main result is that the stationary probabilities themselves satisfy mutual linearity. For two observation points $x_1$ and $x_2$, one has (see \cite{supp} for detailed derivations)
\begin{equation}
  \pi_{\lambda}(x_1) = \chi_{x_1 x_2} \, \pi_{\lambda}(x_2) + \gamma_{x_1 x_2}.
  \label{eq:elementary_ml}
\end{equation}
Both the slope $\chi_{x_1 x_2}$ and the intercept $\gamma_{x_1 x_2}$ are independent of the perturbation strength $\lambda$.

The origin of \cref{eq:elementary_ml} is a simple local response geometry. Define the fundamental kernel $Z$ through the propagator $P(x,t | z,0)$ as
\begin{equation}
  Z_{\lambda}(x | z) \equiv \int_0^{\infty} \bigl[ P_{\lambda}(x,t | z,0) - \pi_{\lambda}(x) \bigr] \rmd t,
\end{equation}
and the local response mode 
\begin{equation}
  v_{\lambda,z_0}(x) \equiv \partial_z Z_{\lambda}(x | z)\big|_{z = z_0}.
\end{equation}
As shown in the Supplemental Material \cite{supp}, the same localized perturbation implies two proportionality relations:
\begin{subequations}
\begin{align}
  \frac{\rmd \pi_{\lambda}(x)}{\rmd \lambda}  &= \ell_{\phi,z_0}[\pi_\lambda] v_{\lambda,z_0}(x), \label{eq:pi_projection} \\
  \frac{\rmd v_{\lambda,z_0}(x)}{\rmd \lambda}  &= \ell_{\phi,z_0}[v_{\lambda,z_0}] v_{\lambda,z_0}(x), \label{eq:v_projection}
\end{align}
\end{subequations}
Here, an infinitesimal perturbation of strength $\epsilon$ at $z_0$ modifies the current operator according to $\mathcal{J} \mapsto \mathcal{J}_{\phi}^{\epsilon,z_0} \equiv \mathcal{J}+\epsilon\,\delta(x-z_0)\mathcal{K}_{\phi}$; the operator $\mathcal{K}_{\phi}$ depends on which parameter is perturbed (e.g. $\mathcal{K}_{F} = \mu$ for force perturbations, see \cite{supp}). We then define $\ell_{\phi,z_0}[f] \equiv (\mathcal{K}_{\phi} f)(z_0)$. Equation \eqref{eq:pi_projection} states that the density moves along a single response mode $v_{\lambda,z_0}(x)$, while \eqref{eq:v_projection} states that the mode $v_{\lambda,z_0}(x)$ does not rotate as $\lambda$ varies. Consequently, the slope $\chi_{x_1 x_2} \equiv \frac{\rmd_\lambda \pi_\lambda(x_1)}{\rmd_\lambda \pi_\lambda(x_2)}$ is independent of $\lambda$, $\rmd \chi_{x_1 x_2}/\rmd \lambda = 0$, which yields \cref{eq:elementary_ml}. Additionally, the slope and intercept can be expressed as
\begin{subequations}
\begin{align}
  \chi_{x_1 x_2} &= \frac{v_{\lambda,z_0}(x_1)}{v_{\lambda,z_0}(x_2)}, \\
  \gamma_{x_1 x_2} &= \pi_{\lambda}(x_1) - \chi_{x_1 x_2} \pi_{\lambda}(x_2).
\end{align}
\end{subequations}

\emph{State-current mutual linearity.}--- We now consider the state-current observables in \cref{eq:observable}. For observables whose current component does not probe the perturbed point, that is, $b(z_0) = 0$, the stationary average can be written as a linear functional of the stationary density,
\begin{equation}
  Q_{\mathrm{ss},\lambda} = \int \rmd x \, c_Q(x)\pi_{\lambda}(x),
  \label{eq:observable_density_functional_main}
\end{equation}
where $c_Q(x) = a(x) + \mathcal{J}_0^{\dagger} b(x)$ is determined by the observable, and $\mathcal{J}_0^{\dagger}$ is the adjoint of the unperturbed current operator $\mathcal{J}_0$. The exclusion of the position $z_0$ ensures that the coefficient $c_Q(x)$ does not depend on $\lambda$. The elementary mutual linearity of the stationary density therefore immediately implies our second main result: for any two excluded-point state-current observables $Q_1$ and $Q_2$ (see Supplemental Material \cite{supp} for detailed derivations),
\begin{equation}
  Q_{1,\mathrm{ss},\lambda} = \chi_{12} \, Q_{2,\mathrm{ss},\lambda} + \gamma_{12}.
  \label{eq:observable_ml_main}
\end{equation}
Thus, once the perturbed position is removed from the observable, the mutual linearity of observables follows directly from the mutual linearity of the stationary density. The slope $\chi_{12}$ and the intercept $\gamma_{12}$ can be expressed as
\begin{subequations}
\begin{align}
  \chi_{12} &= \frac{\int \rmd x \, c_{Q_1}(x) v_{\lambda,z_0}(x)}{\int \rmd x \, c_{Q_2}(x) v_{\lambda,z_0}(x)}, \\
  \gamma_{12} &= Q_{1,\mathrm{ss},\lambda} - \chi_{12} Q_{2,\mathrm{ss},\lambda}.
\end{align}
\end{subequations}

\emph{Non-stationary mutual linearity.}--- Mutual linearity also persists in relaxation dynamics. Let $p_{\lambda}(x,t)$ denote the time-dependent density relaxation under the time-homogeneous dynamics generated by $\phi_{\lambda}(x)$. Assume the initial density $p(x, 0)$ is independent of $\lambda$. Consider the Laplace transform $\hat{p}_{\lambda}(x,\omega) \equiv \int_0^{\infty} e^{-\omega t} p_{\lambda}(x,t) \, \rmd t$. Our third main result is that the Laplace-transformed non-stationary probabilities satisfy (see Supplemental Material \cite{supp} for detailed derivations)
\begin{equation}
  \hat{p}_{\lambda}(x_1,\omega) = \hat{\chi}_{x_1 x_2}(\omega) \, \hat{p}_{\lambda}(x_2,\omega) + \hat{\gamma}_{x_1 x_2}(\omega).
  \label{eq:nonstationary_density_ml}
\end{equation}
Correspondingly, the Laplace transforms of excluded-point state-current observables also satisfy
\begin{equation}
  \hat{Q}_{1,\lambda}(\omega) = \hat{\chi}_{12}(\omega) \, \hat{Q}_{2,\lambda}(\omega) + \hat{\gamma}_{12}(\omega),
  \label{eq:nonstationary_observable_ml}
\end{equation}
where $\hat{Q}_{\lambda}(\omega) \equiv \int_0^{\infty} e^{-\omega t} Q_{\lambda}(t) \, \rmd t$. The slope and intercept are given by
\begin{subequations}
\begin{align}
  \hat{\chi}_{12}(\omega) &= \frac{\int \rmd x \, c_{Q_1}(x) \hat{v}_{\lambda,z_0}(x,\omega)}{\int \rmd x \, c_{Q_2}(x) \hat{v}_{\lambda,z_0}(x,\omega)}, \\
  \hat{\gamma}_{12}(\omega) &= \hat{Q}_{1,\lambda}(\omega) - \hat{\chi}_{12}(\omega) \hat{Q}_{2,\lambda}(\omega),
\end{align}
\end{subequations}
where $\hat v_{\lambda,z_0}(x,\omega)\equiv \partial_z \hat Z_\lambda(x|z;\omega)|_{z=z_0}$ is the local response mode associated with the Laplace-domain resolvent. The derivation is given in the Supplemental Material \cite{supp}. Essentially, even for non-stationary relaxation processes, the projection relations \cref{eq:pi_projection,eq:v_projection} are preserved in the Laplace domain.

\emph{Unified origin of mutual linearity.}--- This letter reveals that mutual linearity has the same structural origin in both discrete and continuous Markov systems. As summarized in \cref{tab:comparison_mjp_langevin}, a local perturbation selects a single response mode $v$: in a Markov network, $v$ is the finite difference of the fundamental matrix along the perturbed edge, whereas in an overdamped Langevin system, $v$ is the spatial derivative of the fundamental kernel at the perturbed point. In both cases, the stationary distribution responds along this mode, $\partial_{\lambda}\pi \propto v$, and the mode itself remains parallel to itself as the perturbation strength is varied, $\partial_{\lambda}v \propto v$. These two relations imply that the ratio between the responses at any two states or positions is independent of the perturbation strength, yielding mutual linearity.

This structure identifies the two essential ingredients of mutual linearity: locality and rigidity. Locality ensures that the perturbation excites a single response direction, while rigidity ensures that this direction is preserved under finite changes of the same perturbation. Therefore, mutual linearity does not rely on the discreteness of the state space or on a special graph topology. The factorized response relation derived in \cite{chun2026fluctuation} corresponds to the infinitesimal local-response structure. The rigidity relation established here further promotes this infinitesimal factorization to a finite affine relation among stationary densities and excluded-point state-current observables.

\begin{table*}[htbp]
  \caption{Comparison of the basic structures underlying mutual linearity in Markov jump processes and overdamped Langevin dynamics.}
  \label{tab:comparison_mjp_langevin}
  \centering
  \begin{tabular}{lll}
    \toprule
    \PCell{0.30\textwidth}{Quantity} & \PCell{0.30\textwidth}{Markov jump process} & \PCell{0.35\textwidth}{overdamped Langevin dynamics} \\
    \midrule
    fundamental matrix/kernel &
    $Z_{ij} = \int_0^{\infty} \bigl[ P(i, t|j, 0) - \pi_i \bigr] \rmd t$ &
    $Z(x | z) = \int_0^{\infty} \bigl[ P(x,t | z,0) - \pi(x) \bigr] \rmd t$ \\[6pt]

    response local mode &
    $v_{k, ij} \equiv Z_{k i} - Z_{k j}$ &
    $v_{\lambda, z_0}(x) \equiv \partial_z Z_{\lambda}(x | z)\big|_{z = z_0}$ \\[6pt]

    steady-state response &
    $\dfrac{\rmd \pi_k}{\rmd r_{ij}} = \pi_j v_{k, ij}$ \cite{zheng2026mutual} &
    $\dfrac{\rmd \pi_{\lambda}(x)}{\rmd \lambda} = \ell_{\phi,z_0}[\pi_\lambda] v_{\lambda,z_0}(x)$ \\[8pt]

    mode response &
    $\dfrac{\rmd v_{k, ij}}{\rmd r_{ij}} = v_{j, ij} v_{k, ij}$ \cite{zheng2026mutual} &
    $\dfrac{\rmd v_{\lambda, z_0}(x)}{\rmd \lambda} = \ell_{\phi,z_0} \!\left[ v_{\lambda, z_0} \right] v_{\lambda, z_0}(x)$ \\
    \bottomrule
  \end{tabular}
\end{table*}

\emph{Finite-width Gaussian perturbations.}---In experiments, an ideal $\delta$-like perturbation is rarely accessible. We therefore consider a finite-width local perturbation:
\begin{subequations}
\begin{align}
  \phi_{\lambda,h}(x) &= \phi_0(x) + \lambda G_h(x-z_0), \\
  G_h(x-z_0) &= \frac{1}{h\sqrt{2\pi}}\exp\!\left[-\frac{(x-z_0)^2}{2h^2}\right],
\end{align}
\label{eq:gaussian_perturbation}
\end{subequations}
where $G_h$ is a normalized Gaussian packet of variance $h^2$. The width of the packet is usually captured by the standard deviation $h$. A finite-width perturbation averages nearby local response modes. Since the Gaussian has zero first moment, the leading correction is controlled by its variance. For any smooth test function $f$,
\begin{equation}
  \int \rmd x\,G_h(x-z_0)f(x) = f(z_0) + \frac{h^2}{2}f''(z_0) + O(h^4).
  \label{eq:gaussian_moment_expansion}
\end{equation}
Thus, the finite-width perturbation differs from the ideal local perturbation only at order $O(h^2)$.

Consequently, the exact affine relations become asymptotically accurate for sufficiently narrow Gaussian packet perturbations. For any pair of density values or excluded-point state-current observables, denoted collectively by $Q_1$ and $Q_2$, one obtains
\begin{equation}
  Q_{1,\lambda}^{(h)} = \chi_{12}Q_{2,\lambda}^{(h)} + \gamma_{12} + h^2\Delta_{12}(\lambda) + O(h^4).
  \label{eq:gaussian_affine_error}
\end{equation}
The explicit expression for the coefficient $\Delta_{12}(\lambda)$ is given in the Supplemental Material \cite{supp}. For current-like observables, the excluded-point condition should be understood on the scale of the packet width: the current weight $b(x)$ should locally vanish, or be negligible, in the perturbation region where $G_h(x-z_0)$ has appreciable weight. If the quality of the linear relation is quantified by the coefficient of determination $R^2$, then we have $1-R^2=O(h^4)$. Thus, the affine residual is $O(h^2)$, while $1 - R^2 = O(h^4)$, showing that mutual linearity is robust to sufficiently localized finite-width perturbations.

\emph{F$_1$-ATPase.}--- As a concrete illustration, we consider the rotary molecular motor F$_1$-ATPase under external torque. Following the standard tilted-periodic-potential description of the bead-coupled motor \cite{reimann2001giant,toyabe2011thermodynamic,mishima2025efficiently}, we model the unwrapped angular coordinate $\theta_t$ by
\begin{equation}
  \dot{\theta}_t = \mu \Bigl[ -\partial_{\theta} U(\theta)\big|_{\theta = \theta_t} + F_{\mathrm{drive}} + \lambda G_{h}(\theta_t-z_0) \Bigr] + \sqrt{2\mu T} \, \xi_t,
  \label{eq:f1_model}
\end{equation}
with $U(\theta)=U_0\cos(3\theta)$. Here $F_{\mathrm{drive}}$ is the externally applied torque, and $G_{h}$ is a normalized $2\pi$-periodic Gaussian torque packet centered at the barrier $z_0=0$. The width $h$ controls how localized the perturbation is. We compare three cases: a narrow packet, $h=0.05\,\mathrm{rad}$; an intermediate-width packet, $h=0.5\,\mathrm{rad}$; and a broad packet, $h=1.0\,\mathrm{rad}$.

The simulation parameters are chosen to match the experimentally relevant F$_1$-ATPase regime near giant diffusion \cite{reimann2001giant}. We list them under physical units: $T=298\,\mathrm{K}$, $\mu=0.91\,\mathrm{rad}/(\mathrm{s}\,\mathrm{pN}\,\mathrm{nm})$, $U_0=10k_{\mathrm B}T$, and $F_{\mathrm{drive}}=120\,\mathrm{pN}\,\mathrm{nm}$. Numerical details are given in the End Matter. The dynamics is integrated with an overdamped Euler--Maruyama scheme using $\rmd t=10^{-5}\,\mathrm{s}$. For each value of $\lambda$, we simulate $5\times 10^4$ independent trajectories in parallel.

We monitor several state and state-current observables. First, we divide the angular coordinate into three non-overlapping sectors centered at the three minima $\theta_A=\pi/3$, $\theta_B=\pi$, and $\theta_C=5\pi/3$, and measure their occupation fractions
\begin{equation}
  \tau_{m,\mathrm{ss}} = \int \rmd \theta \, W_m(\theta)\pi_{\lambda}(\theta), \qquad m\in\{A,B,C\}.
  \label{eq:f1_sector_occupations}
\end{equation}
Here $W_m(\theta)$ is the indicator window for sector $m$, with $W_A+W_B+W_C=1$. Second, we estimate the stationary density at $\theta_A$ and $\theta_B$ using narrow Gaussian probes, denoted by $\tilde{\pi}_{\rho}(\theta_A)$ and $\tilde{\pi}_{\rho}(\theta_B)$. Third, we measure two current-type observables localized in sector $B$: (i) the local input-power observable
\begin{equation}
  P_B^{\mathrm{in}} = F_{\mathrm{drive}}\int \rmd \theta \, \widetilde{W}_B(\theta)j_{\mathrm{ss},\lambda}(\theta),
  \label{eq:f1_power_observable}
\end{equation}
and (ii) the local entropy-production rate
\begin{equation}
  \sigma_B = \frac{1}{T} \int \rmd \theta \, \widetilde{W}_B(\theta) F_{\lambda}^{\mathrm{tot}}(\theta) j_{\mathrm{ss},\lambda}(\theta),
  \label{eq:f1_local_entropy_production}
\end{equation}
where $F_{\lambda}^{\mathrm{tot}}(\theta) \equiv -\partial_{\theta}U(\theta)+F_{\mathrm{drive}}+\lambda G_h(\theta-z_0)$
is the total torque acting on the angular coordinate. The same smooth window $W_B(\theta)$ is used for both current-type observables:
\begin{equation}
  \widetilde{W}_B(\theta) = \frac{1}{2} \left[ 1+ \tanh\!\left( \frac{h_P-\left|d_{2\pi}(\theta,\theta_B)\right|}{e_P} \right) \right],
  \label{eq:f1_power_window}
\end{equation}
where $d_{2\pi}(\theta,\theta_B)$ is the shortest periodic distance between $\theta$ and $\theta_B$, and we use $h_P=0.65\,\mathrm{rad}$ and $e_P=0.06\,\mathrm{rad}$. This window localizes the current measurements to sector $B$, away from the perturbed barrier at $z_0=0$. Therefore, both $P_B^{\mathrm{in}}$ and $\sigma_B$ are excluded-point state-current observables of the type covered by \cref{eq:observable_ml_main}.

\begin{figure}[htbp]
  \centering
  \includegraphics[width=0.95\linewidth]{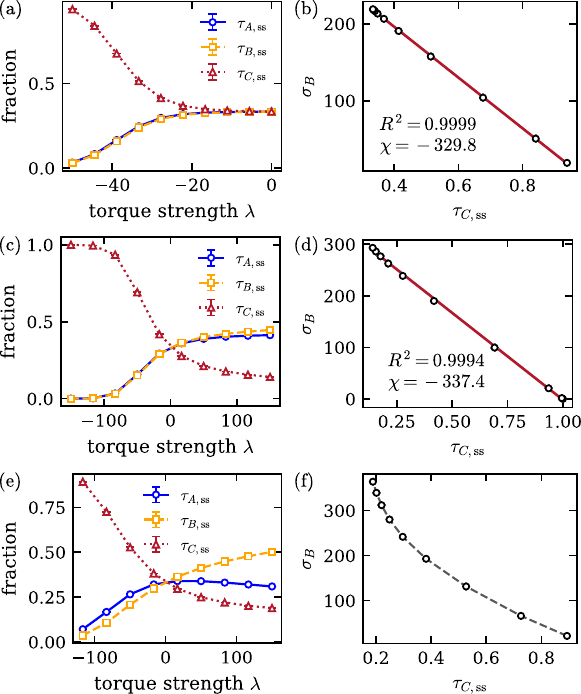}
  \caption{Steady-state mutual linearity and its breakdown in the F$_1$-ATPase model. Panels (a,b), (c,d), and (e,f) correspond to Gaussian widths $h=0.05$, $0.5$, and $1.0\,\mathrm{rad}$, respectively. Panels (a), (c), and (e) show the nonlinear dependence of representative observables on the perturbation strength $\lambda$. Panels (b) and (d) show the mutual-linearity relation between the local entropy-production rate $\sigma_B$ and the sector occupation $\tau_{C,\mathrm{ss}}$. Notice as perturbation delocalize, i.e., $h=1.0$, the mutual-linearity is broken in (f).}
  \label{fig:ness_ml}
\end{figure}

The steady-state results are shown in \cref{fig:ness_ml}. For each perturbation width, we show one representative nonlinear response to the perturbation strength $\lambda$ and one mutual-linearity plot between the local entropy-production rate $\sigma_B$ and the sector occupation $\tau_{C,\mathrm{ss}}$. For the narrow packet $h=0.05\,\mathrm{rad}$, $\sigma_B$ and $\tau_{C,\mathrm{ss}}$ collapse onto a straight line, supporting the local-perturbation prediction. For the intermediate packet $h=0.5\,\mathrm{rad}$, the affine relation remains accurate, demonstrating the finite-width robustness discussed above. For the broad packet $h=1.0\,\mathrm{rad}$, the relation becomes visibly nonlinear, showing that sufficiently broad perturbations destroy the single-mode response structure. Full steady-state mutual-linearity tests for additional observable pairs are given in the End Matter.

\begin{figure}[htbp]
  \centering
  \includegraphics[width=0.95\linewidth]{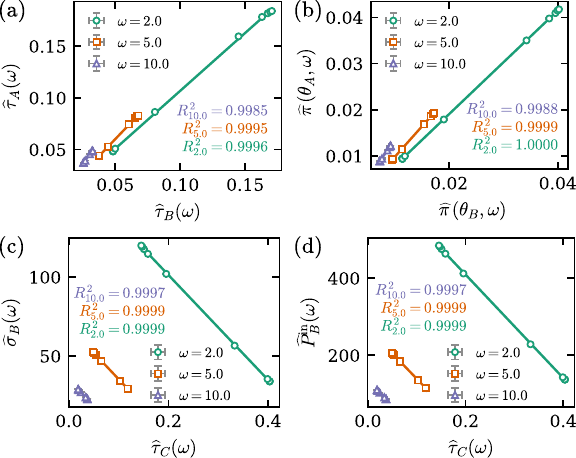}
  \caption{Non-stationary mutual linearity in the Laplace domain. The Gaussian width is fixed at $h=0.1\,\mathrm{rad}$, and the same $\lambda$-independent initial distribution is used for all perturbation strengths. The plotted quantities are truncated Laplace transforms of the time-dependent observables, evaluated at $\omega=2.0$, $5.0$, and $10.0$. The panels show mutual-linearity plots between occupation fractions, density probes, local input power, and local entropy production.}
  \label{fig:nonstationary_ml}
\end{figure}

We further test the non-stationary prediction in the Laplace domain, as shown in \cref{fig:nonstationary_ml}. For a fixed width $h=0.1\,\mathrm{rad}$, we illustrate the mutual linearity at frequencies $\omega = 2.0$, $5.0$, and $10.0$. This agreement supports the theoretical result that non-stationary relaxation dynamics inherit the same one-dimensional response structure as that of the steady states.

\emph{Conclusions and outlook.}--- This Letter shows that mutual linearity is not a special feature of discrete Markov networks, but a geometric response structure that also appears in continuous nonequilibrium dynamics. In overdamped Langevin systems, a localized perturbation selects a single response mode, and this mode remains parallel to itself as the perturbation strength is varied. This one-dimensional response geometry explains why stationary densities, excluded-point state-current observables, and their Laplace-domain non-stationary counterparts satisfy affine relations. The finite-width analysis and the F$_1$-ATPase simulations further show that the predicted structure is not destroyed by experimentally realistic perturbations, as long as the perturbation remains sufficiently localized.

Beyond its intellectual value, mutual linearity suggests a practical inference strategy for experiments. Many complex nonequilibrium observables, such as local entropy production or hidden dissipation, are difficult to measure directly, whereas occupations, local densities, or coarse-grained currents are often more accessible. Once the affine relation between an accessible observable and a target observable is calibrated using two perturbation strengths, later measurements of the accessible observable can be used to infer the desired inaccessible observables. This provides a possible route to probe response and dissipation in molecular motors, biophysical transport, active matter, and driven soft-matter systems. In the End Matter, we propose a concrete experimental setup to test this prediction.

\emph{Acknowledgements.}--- This work is supported by the U.S. National Science Foundation under Grant No. DMR-2145256 and Alfred P. Sloan Foundation Matter-to-Life Theory Award under Grant No. G-2025-25194.

\emph{Data availability.}--- The data that support the findings of this article are generated by numerical simulation codes that are openly available at \cite{data}.

\bibliographystyle{apsrev4-2}
\bibliography{manuscript}

\section{End Matter}

\emph{Detailed steady-state simulations.}---
In the main text, we showed a representative steady-state mutual-linearity test using the local entropy-production observable $\sigma_B$. Here we provide the complete set of steady-state simulation results for the F$_1$-ATPase model. The same three perturbation widths are considered: $h=0.05$, $0.5$, and $1.0\,\mathrm{rad}$. For each width, we plot one representative nonlinear response to the perturbation strength $\lambda$, followed by four mutual-linearity plots involving sector occupations, smoothed density probes, the local entropy-production rate, and the local input-power observable.

\begin{figure*}[htbp]
  \centering
  \includegraphics[width=\linewidth]{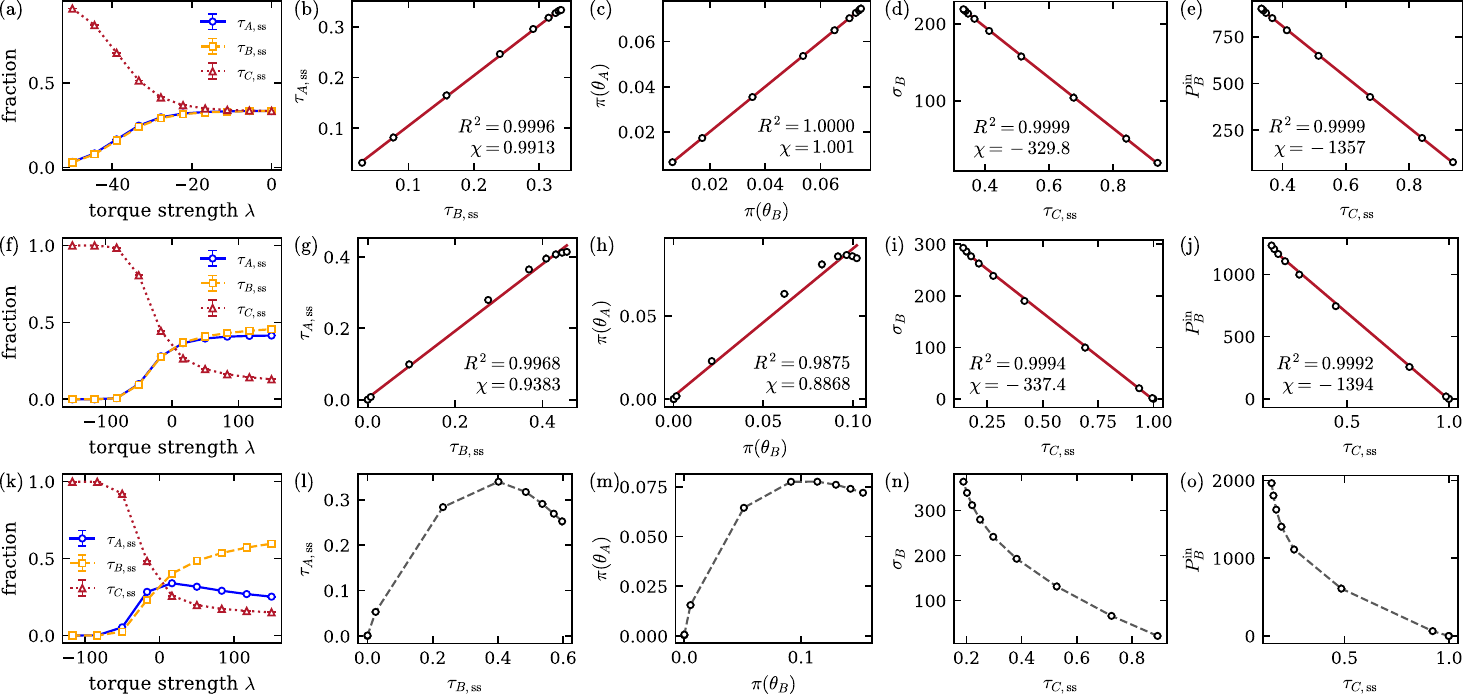}
  \caption{Complete steady-state mutual-linearity tests for the F$_1$-ATPase model. Panels (a--e), (f--j), and (k--o) correspond to Gaussian widths $h=0.05$, $0.5$, and $1.0\,\mathrm{rad}$, respectively. Panels (a), (f), and (k) show representative observables as functions of the perturbation strength $\lambda$, illustrating that individual responses are generally nonlinear. The remaining panels show mutual-linearity plots between occupation fractions, smoothed density probes, the local entropy-production observable, and the local input-power observable: $\tau_{A,\mathrm{ss}}$ versus $\tau_{B,\mathrm{ss}}$, $\tilde{\pi}_{\rho}(\theta_A)$ versus $\tilde{\pi}_{\rho}(\theta_B)$, $\sigma_B$ versus $\tau_{C,\mathrm{ss}}$, and $P_B^{\mathrm{in}}$ versus $\tau_{C,\mathrm{ss}}$. The narrow and intermediate-width perturbations preserve strong affine relations, whereas the broad perturbation visibly breaks mutual linearity.}
  \label{fig:ness_full}
\end{figure*}

As shown in \cref{fig:ness_full}, the narrow perturbation $h=0.05\,\mathrm{rad}$ gives essentially perfect affine collapse for all tested observable pairs, consistent with the ideal local-perturbation theory. For the intermediate width $h=0.5\,\mathrm{rad}$, the perturbation is no longer sharply localized, but the mutual-linearity plots remain close to straight lines for all observable pairs, supporting the predicted robustness of mutual linearity under finite-width perturbations. For the broad perturbation $h=1.0\,\mathrm{rad}$, the observable pairs develop pronounced nonlinear curves. This breakdown is consistent with the finite-width analysis: a broad perturbation averages over nearby local response modes and no longer selects a single one-dimensional response direction.

\emph{Proposed experimental test.---}A direct test of the predicted mutual linearity can be implemented with a single colloidal particle confined to a ring-shaped optical or feedback trap. Such toroidal-trap platforms have already been used to realize overdamped Brownian motion on a circle and to generate nonequilibrium steady states by combining a periodic optical potential with a nonconservative tangential drive \cite{blickle2007einstein,blickle2007characterizing,gomez2009experimental,blickle2009relaxation}. In this setting, the angular coordinate $\theta$ of the particle provides a controllable one-dimensional Langevin degree of freedom on a periodic domain. The unperturbed dynamics can be prepared by imposing a baseline potential $U_0(\theta)$ together with a constant tangential drive, producing a nonequilibrium steady state with nonzero stationary current.

The local perturbation required by our theory can be implemented by real-time feedback control. Optical feedback traps have been shown to synthesize prescribed virtual potentials and force landscapes by measuring the particle position and updating the trap position or force accordingly \cite{jun2012virtual,gavrilov2017feedback,kumar2018nanoscale,albay2018optical}. Using this approach, one can add an angle-dependent tangential force of the form $\lambda G_h(\theta-z_0)$, where $G_h$ is a narrow periodic Gaussian packet centered at a chosen angular position $z_0$. The perturbation strength $\lambda$ is then scanned while keeping $z_0$ and $h$ fixed. Dynamic or holographic optical tweezer techniques, which have been used to implement time-dependent colloidal force landscapes and reconstruct probability currents in nonequilibrium systems \cite{curtis2002dynamic,grier2006holographic,roichman2007colloidal,thapa2024nonequilibrium}, provide an experimentally viable route for applying such controlled local force packets and measuring the resulting steady-state trajectories.

For each value of $\lambda$, long trajectories can be used to reconstruct the stationary density $\pi_\lambda(\theta)$, the stationary current $j_{{\rm ss},\lambda}(\theta)$, sector occupation fractions, local density probes, and windowed current observables away from the perturbed region. The characteristic experimental signature is not that each observable depends linearly on $\lambda$. Instead, individual observables may vary nonlinearly with $\lambda$, while pairwise plots between any two excluded-point state-current observables should collapse onto affine lines,
\begin{equation}
  Q_{1,{\rm ss},\lambda} = \chi_{12}Q_{2,{\rm ss},\lambda} + \gamma_{12},
\end{equation}
with $\chi_{12}$ and $\gamma_{12}$ independent of $\lambda$. For example, one may compare two sector occupations, two local density probes, or a windowed current observable and a sector occupation, provided that the corresponding measurement windows do not overlap with the region where $G_h(\theta-z_0)$ has appreciable weight.

The same platform also allows a direct test of the finite-width prediction. By repeating the experiment for several packet widths $h$, one can quantify the residual from the best affine fit,
\begin{equation}
  \epsilon_\lambda^{(h)} = Q_{1,{\rm ss},\lambda}^{(h)} - \chi_{12}^{(h)} Q_{2,{\rm ss},\lambda}^{(h)} - \gamma_{12}^{(h)} .
\end{equation}
In the narrow-packet regime, our theory predicts that the affine residual $\epsilon_\lambda^{(h)}$ scales as $O(h^2)$, or equivalently that $1-R^2$ scales as $O(h^4)$. A useful control experiment is to replace the single local packet by two separated packets or to choose a current observable whose window overlaps the perturbed region; in both cases the single-mode response structure is expected to be degraded, leading to visible deviations from mutual linearity.

\end{document}

\preprint{APS/123-QED}

\title{Supplemental Material for ``Mutual Linearity in Nonequilibrium Langevin Dynamics''}
\author{Jiming Zheng} \email{jiming@unc.edu}
\author{Zhiyue Lu} \email{zhiyuelu@unc.edu}
\affiliation{Department of Chemistry, University of North Carolina-Chapel Hill, NC}

\date{\today}

\maketitle

In this Supplemental Material, we provide complementary derivations of mutual linearity in overdamped Langevin dynamics. The derivation is organized as follows. In \cref{SMsec:response}, we derive the response of a general observable to a local perturbation in terms of the fundamental kernel. In \cref{SMsec:local-ratio}, we show how the one-kernel structure of the local response leads to steady-state mutual linearity. In \cref{SMsec:laplace}, we show that the same structure persists after Laplace transformation, yielding frequency-domain mutual linearity. In \cref{SMsec:multi}, we extend the same rank-one structure to multidimensional overdamped Langevin dynamics, both in the steady state and in the Laplace domain. In \cref{SMsec:finite-width}, we provide detailed derivations of the finite-width correction effect. In \cref{SMsec: numerical}, we provide detailed information about numerical simulations in the main text.

\tableofcontents

\newpage

\section{Response Theory for Overdamped Langevin Dynamics}
\label{SMsec:response}

\subsection{Overdamped Langevin dynamics}

We consider a one-dimensional overdamped Langevin system described by
\begin{equation}
  \dot{x}_t = \mu(x_t)F(x_t) + \sqrt{2\mu(x_t)T(x_t)} \,\star\, \xi_t .
  \label{SMeq:langevin}
\end{equation}
Here, $x_t$ is the stochastic state variable, $\mu(x)$ is the position-dependent mobility, $F(x)$ is the systematic force, and $T(x)$ is the local temperature field. The symbol $\star$ indicates the anti-It\^o convention required for thermodynamic consistency \cite{lau2007state}. The noise $\xi_t$ is a Gaussian white noise with zero mean and correlation function $\langle \xi_t\xi_{t'}\rangle = \delta(t-t')$. Throughout this work, we set the Boltzmann constant to unity.

The probability density $p(x,t)$ associated with \cref{SMeq:langevin} obeys the Fokker--Planck equation
\begin{equation}
  \partial_t p(x,t) = -\partial_x \! \left[ \mu(x)F(x)p(x,t) - \mu(x)T(x)\partial_x p(x,t) \right].
  \label{SMeq:fp}
\end{equation}
It is convenient to introduce the probability-current operator
\begin{equation}
  \mathcal{J} = \mu(x)\bigl[F(x)-T(x)\partial_x\bigr],
  \label{SMeq:current_operator}
\end{equation}
and the Fokker--Planck generator
\begin{equation}
  \mathcal{L} = -\partial_x \mathcal{J},
  \label{SMeq:generator}
\end{equation}
so that \cref{SMeq:fp} can be written compactly as $\partial_t p(x,t)=\mathcal{L}p(x,t)$. We assume that the dynamics admits a unique stationary density $\pi(x)$ satisfying $\mathcal{L}\pi(x)=0$, and denote the corresponding stationary current by $j_{\mathrm{ss}}(x)=\mathcal{J}\pi(x)$.

We focus on state-current observables of the form
\begin{equation}
  Q(t) = \int dx\,a(x)p(x,t) + \int dx\,b(x)j(x,t),
  \label{SMeq:observable}
\end{equation}
where $a(x)$ and $b(x)$ are arbitrary smooth functions. The first term represents the state observable, while the second term represents the current observable. In the steady state, $p(x,t)=\pi(x)$ and $j(x,t)=j_{\mathrm{ss}}(x)$, so \cref{SMeq:observable} reduces to
\begin{equation}
  Q_{\mathrm{ss}} = \int dx\,a(x)\pi(x) + \int dx\,b(x)j_{\mathrm{ss}}(x).
  \label{SMeq:observable_ss}
\end{equation}

\subsection{Response to local perturbations expressed through fundamental kernel}
\label{SMsec:response_local}

We now derive the response of a stationary observable to a local perturbation. We also emphysize that the response relations has been derived in \cite{chun2026fluctuation}. In the present continuous setting, the natural analogue of a local perturbation is to perturb one of the dynamical parameters $\phi(x) \in \{F(x),\mu(x),T(x)\}$ at a fixed position $z$ as
\begin{equation}
  \phi(x) \rightarrow \phi(x) + \epsilon\,\delta(x-z).
  \label{SMeq:local_parameter_perturbation}
\end{equation}
For mobility and temperature perturbations, the perturbation amplitude is restricted so that the corresponding regularized $\mu(x)$ and $T(x)$ remain positive.

Since the probability-current operator
\begin{equation}
  \mathcal{J} = \mu(x)\bigl[F(x)-T(x)\partial_x\bigr]
  \label{SMeq:current_operator_recalled}
\end{equation}
depends explicitly on $F(x)$, $\mu(x)$, and $T(x)$, such a local perturbation induces a corresponding local deformation of $\mathcal{J}$. The perturbed current operator can be written as
\begin{equation}
  \mathcal{J}_{\phi}^{\epsilon,z} \equiv \mathcal{J} + \epsilon\,\delta(x-z)\mathcal{K}_{\phi},
  \label{SMeq:perturbed_current_operator}
\end{equation}
where $\mathcal{K}_{\phi}$ specifies the perturbed local channel. The explicit form of $\mathcal{K}_{\phi}$ follows from a first-order expansion of $\mathcal{J}$ with respect to the perturbed parameter.

For a local perturbation of the force field,
\begin{equation}
  F(x) \rightarrow F(x) + \epsilon\,\delta(x-z),
  \label{SMeq:force_perturbation}
\end{equation}
the current operator becomes
\begin{equation}
  \mathcal{J}_{F}^{\epsilon,z} = \mu(x)\bigl[F(x)+\epsilon\,\delta(x-z)-T(x)\partial_x\bigr] = \mathcal{J}+\epsilon\,\delta(x-z)\mu(x),
  \label{SMeq:J_force_perturbation}
\end{equation}
so that
\begin{equation}
  \mathcal{K}_F \equiv \mu(x).
  \label{SMeq:KF}
\end{equation}

For a local perturbation of the mobility,
\begin{equation}
  \mu(x) \rightarrow \mu(x) + \epsilon\,\delta(x-z),
  \label{SMeq:mobility_perturbation}
\end{equation}
we obtain
\begin{equation}
  \mathcal{J}_{\mu}^{\epsilon,z} = \bigl[\mu(x)+\epsilon\,\delta(x-z)\bigr]\bigl[F(x)-T(x)\partial_x\bigr] = \mathcal{J}+\epsilon\,\delta(x-z)\bigl[F(x)-T(x)\partial_x\bigr],
  \label{SMeq:J_mobility_perturbation}
\end{equation}
hence
\begin{equation}
  \mathcal{K}_{\mu} \equiv F(x)-T(x)\partial_x.
  \label{SMeq:Kmu}
\end{equation}

For a local perturbation of the temperature,
\begin{equation}
  T(x) \rightarrow T(x) + \epsilon\,\delta(x-z),
  \label{SMeq:temperature_perturbation}
\end{equation}
the current operator becomes
\begin{equation}
  \mathcal{J}_{T}^{\epsilon,z} = \mu(x)\bigl[F(x)-\bigl(T(x)+\epsilon\,\delta(x-z)\bigr)\partial_x\bigr] = \mathcal{J}-\epsilon\,\delta(x-z)\mu(x)\partial_x,
  \label{SMeq:J_temperature_perturbation}
\end{equation}
so that
\begin{equation}
  \mathcal{K}_T \equiv -\mu(x)\partial_x.
  \label{SMeq:KT}
\end{equation}

The corresponding perturbed generator is
\begin{equation}
  \mathcal{L}_{\phi}^{\epsilon,z} \equiv -\partial_x \mathcal{J}_{\phi}^{\epsilon,z}.
  \label{SMeq:perturbed_generator}
\end{equation}

Let $P(x,t|z,0)$ be the propagator generated by the unperturbed generator $\mathcal{L}$. We define the fundamental kernel
\begin{equation}
  Z(x|z) \equiv \int_0^\infty \bigl[P(x,t|z,0)-\pi(x)\bigr]dt.
  \label{SMeq:fundamental_kernel}
\end{equation}
By construction, it satisfies
\begin{equation}
  \mathcal{L}_x Z(x|z)=\pi(x)-\delta(x-z),
  \label{SMeq:fundamental_kernel_equation}
\end{equation}
where the subscript on $\mathcal{L}_x$ indicates that the generator acts on the $x$ variable. Since both $P(x,t|z,0)$ and $\pi(x)$ are normalized in $x$, the fundamental kernel also obeys
\begin{equation}
  \int dx\,Z(x|z)=0.
  \label{SMeq:fundamental_kernel_zero_mass}
\end{equation}

We next define the local response mode
\begin{equation}
  v_z(x) \equiv \partial_z Z(x|z).
  \label{SMeq:vz_definition}
\end{equation}
Differentiating \cref{SMeq:fundamental_kernel_equation} with respect to $z$, we obtain
\begin{equation}
  \mathcal{L}v_z(x) = -u_z(x),
  \qquad
  u_z(x) \equiv -\partial_x\delta(x-z).
  \label{SMeq:vz_equation}
\end{equation}
Differentiating \cref{SMeq:fundamental_kernel_zero_mass} with respect to $z$ also gives
\begin{equation}
  \int dx\,v_z(x)=0.
  \label{SMeq:vz_zero_mass}
\end{equation}
Thus $v_z(x)$ is the unique zero-mass solution of \cref{SMeq:vz_equation}. This is the common propagating object that controls all local responses.

We now derive the response of the stationary density. For a local perturbation of the parameter $\phi$ at position $z$, we expand the stationary density as
\begin{equation}
  \pi_{\phi}^{\epsilon,z}(x) = \pi(x)+\epsilon\,q_{\phi}(x;z)+O(\epsilon^2).
  \label{SMeq:pi_epsilon_expansion}
\end{equation}
Substituting \cref{SMeq:pi_epsilon_expansion} into the stationary equation $\mathcal{L}_{\phi}^{\epsilon,z}\pi_{\phi}^{\epsilon,z}=0$ and keeping only the terms linear in $\epsilon$, we find
\begin{equation}
  \mathcal{L}q_{\phi}(x;z) = \partial_x\!\left[\delta(x-z)\mathcal{K}_{\phi}\pi\right].
  \label{SMeq:q_equation_raw}
\end{equation}
Using the distributional identity $\delta(x-z)f(x)=f(z)\delta(x-z)$, \cref{SMeq:q_equation_raw} becomes
\begin{equation}
  \mathcal{L}q_{\phi}(x;z) = -N_{\phi}(z)u_z(x),
  \qquad
  N_{\phi}(z) \equiv \bigl(\mathcal{K}_{\phi}\pi\bigr)(z).
  \label{SMeq:q_equation}
\end{equation}
Comparing \cref{SMeq:q_equation} with \cref{SMeq:vz_equation}, we immediately obtain
\begin{equation}
  \frac{\delta \pi(x)}{\delta \phi(z)} = N_{\phi}(z)v_z(x).
  \label{SMeq:density_response_functional}
\end{equation}
For the three perturbation channels, the corresponding prefactors are
\begin{subequations}
\begin{align}
  N_F(z) &= \mu(z)\pi(z), \label{SMeq:NF} \\
  N_{\mu}(z) &= \bigl[F(z)-T(z)\partial_z\bigr]\pi(z) = \frac{j_{\mathrm{ss}}(z)}{\mu(z)}, \label{SMeq:Nmu} \\
  N_T(z) &= -\mu(z)\partial_z\pi(z). \label{SMeq:NT}
\end{align}
\end{subequations}
Equation \cref{SMeq:density_response_functional} already shows the basic factorization structure: the dependence on the perturbed parameter enters only through the scalar prefactor $N_{\phi}(z)$, while the propagation in state space is entirely encoded in the common kernel $v_z(x)$.

We next derive the response of the stationary current. Since $j_{\mathrm{ss}}(x) = \mathcal{J}\pi(x)$, the perturbed stationary current takes the form
\begin{equation}
  j_{\mathrm{ss},\phi}^{\epsilon,z}(x) = \bigl[\mathcal{J}+\epsilon\,\delta(x-z)\mathcal{K}_{\phi}\bigr]\bigl[\pi(x)+\epsilon\,q_{\phi}(x;z)\bigr] + O(\epsilon^2).
  \label{SMeq:current_epsilon_expansion}
\end{equation}
Therefore,
\begin{equation}
  \frac{\delta j_{\mathrm{ss}}(x)}{\delta \phi(z)} = N_{\phi}(z)\bigl[\delta(x-z)+\mathcal{J}v_z(x)\bigr].
  \label{SMeq:current_response_functional}
\end{equation}
The first term in \cref{SMeq:current_response_functional} is a contact term originating from the direct local perturbation of the current operator, while the second term is the propagated contribution through the fundamental kernel.

Using \cref{SMeq:observable_ss,SMeq:density_response_functional,SMeq:current_response_functional}, we obtain the response of the stationary value of the observable $Q$,
\begin{equation}
  \frac{\delta Q_{\mathrm{ss}}}{\delta \phi(z)} = N_{\phi}(z)\left[ b(z) + \int dx\,a(x)v_z(x) + \int dx\,b(x)\mathcal{J}v_z(x) \right].
  \label{SMeq:response_before_adjoint}
\end{equation}
To rewrite the last term in a more convenient form, we introduce the adjoint current operator
\begin{equation}
  \mathcal{J}^{\dagger} b(x) \equiv \mu(x)F(x)b(x) + \partial_x\bigl[\mu(x)T(x)b(x)\bigr].
  \label{SMeq:current_operator_adjoint}
\end{equation}
Under boundary conditions for which the integration-by-parts boundary term vanishes, one has
\begin{equation}
  \int dx\,b(x)\mathcal{J}f(x) = \int dx\,\bigl[\mathcal{J}^{\dagger}b(x)\bigr]f(x).
  \label{SMeq:adjoint_identity}
\end{equation}
Applying \cref{SMeq:adjoint_identity} to \cref{SMeq:response_before_adjoint}, we obtain
\begin{equation}
  \frac{\delta Q_{\mathrm{ss}}}{\delta \phi(z)} = N_{\phi}(z)\left[ b(z) + \int dx\,c_Q(x)v_z(x) \right],
  \qquad
  c_Q(x) \equiv a(x)+\mathcal{J}^{\dagger}b(x).
  \label{SMeq:response_functional_final}
\end{equation}
Thus, if the observable excludes the perturbed point in the sense that
\begin{equation}
  b(z)=0,
  \label{SMeq:excluded_point_condition_local}
\end{equation}
then the contact term disappears, and the response simplifies to
\begin{equation}
  \frac{\delta Q_{\mathrm{ss}}}{\delta \phi(z)} = N_{\phi}(z)\int dx\,c_Q(x)v_z(x).
  \label{SMeq:response_functional_excluded}
\end{equation}
This is the basic one-kernel structure that underlies mutual linearity.

\section{Mutual Linearity for Same Local Perturbation}
\label{SMsec:local-ratio}

In this section, we consider a one-parameter family of localized perturbations applied at a fixed position $z_0$. Our goal is to show that such a perturbation generates a one-dimensional response manifold in the space of stationary states. As a consequence, both the stationary density itself and all excluded-point state-current observables become affinely related to one another as the perturbation strength is varied.

We therefore fix a perturbation channel $\phi \in \{F,\mu,T\}$ and define
\begin{equation}
  \phi_{\lambda}(x) \equiv \phi_{0}(x) + \lambda\,\delta(x-z_0).
  \label{SMeq:philambda}
\end{equation}
The corresponding probability-current operator and generator are
\begin{equation}
  \mathcal{J}_{\lambda} = \mathcal{J}_{0} + \lambda\,\delta(x-z_0)\mathcal{K}_{\phi},
  \qquad
  \mathcal{L}_{\lambda} = -\partial_x\mathcal{J}_{\lambda}.
  \label{SMeq:Jlambda_Llambda}
\end{equation}
For each $\lambda$, we assume that the stationary density $\pi_{\lambda}(x)$ exists and is unique, so that
\begin{equation}
  \mathcal{L}_{\lambda}\pi_{\lambda}(x)=0,
  \qquad
  \int dx\,\pi_{\lambda}(x)=1.
  \label{SMeq:stationary_density_lambda}
\end{equation}
We also denote the corresponding stationary current by
\begin{equation}
  j_{\mathrm{ss},\lambda}(x) \equiv \mathcal{J}_{\lambda}\pi_{\lambda}(x).
  \label{SMeq:stationary_current_lambda}
\end{equation}

\subsection{Mutual linearity of stationary densities}
\label{SMsec:distribution_mutual_linearity}

For each $\lambda$, let $P_{\lambda}(x,t|y,0)$ be the propagator generated by $\mathcal{L}_{\lambda}$, and define the associated fundamental kernel
\begin{equation}
  Z_{\lambda}(x|y) \equiv \int_0^{\infty} \bigl[P_{\lambda}(x,t|y,0)-\pi_{\lambda}(x)\bigr]dt.
  \label{SMeq:fundamental_kernel_lambda}
\end{equation}
By the same argument as in \cref{SMsec:response_local}, $Z_{\lambda}(x|y)$ satisfies
\begin{equation}
  \mathcal{L}_{\lambda,x}Z_{\lambda}(x|y)=\pi_{\lambda}(x)-\delta(x-y),
  \label{SMeq:fundamental_kernel_lambda_equation}
\end{equation}
together with
\begin{equation}
  \int dx\,Z_{\lambda}(x|y)=0.
  \label{SMeq:fundamental_kernel_lambda_zero_mass}
\end{equation}

We now define the local response mode at the perturbed point $z_0$,
\begin{equation}
  v_{\lambda}(x) \equiv \partial_z Z_{\lambda}(x|z)\big|_{z=z_0}.
  \label{SMeq:vlambda_definition}
\end{equation}
Differentiating \cref{SMeq:fundamental_kernel_lambda_equation} with respect to $z$ and setting $z=z_0$, we obtain
\begin{equation}
  \mathcal{L}_{\lambda}v_{\lambda}(x) = -u_{z_0}(x),
  \qquad
  u_{z_0}(x) \equiv -\partial_x\delta(x-z_0).
  \label{SMeq:vlambda_equation}
\end{equation}
Differentiating \cref{SMeq:fundamental_kernel_lambda_zero_mass} with respect to $z$ likewise gives
\begin{equation}
  \int dx\,v_{\lambda}(x)=0.
  \label{SMeq:vlambda_zero_mass}
\end{equation}

Next, we compute the derivative of the generator with respect to $\lambda$. For any test function $f(x)$,
\begin{equation}
  \frac{d\mathcal{L}_{\lambda}}{d\lambda}f = -\partial_x\bigl[\delta(x-z_0)\mathcal{K}_{\phi}f\bigr].
  \label{SMeq:generator_lambda_derivative_raw}
\end{equation}
Using $\delta(x-z_0)g(x)=g(z_0)\delta(x-z_0)$, we can rewrite \cref{SMeq:generator_lambda_derivative_raw} as
\begin{equation}
  \frac{d\mathcal{L}_{\lambda}}{d\lambda}f = u_{z_0}\,\ell_{z_0}[f],
  \qquad
  \ell_{z_0}[f] \equiv \bigl(\mathcal{K}_{\phi}f\bigr)(z_0).
  \label{SMeq:generator_lambda_derivative}
\end{equation}

We first derive the response of the stationary density. Differentiating \cref{SMeq:stationary_density_lambda} with respect to $\lambda$, we find
\begin{equation}
  \frac{d\mathcal{L}_{\lambda}}{d\lambda}\pi_{\lambda} + \mathcal{L}_{\lambda}\frac{d\pi_{\lambda}}{d\lambda} = 0.
  \label{SMeq:pi_lambda_derivative_1}
\end{equation}
Using \cref{SMeq:generator_lambda_derivative}, this becomes
\begin{equation}
  \mathcal{L}_{\lambda}\frac{d\pi_{\lambda}}{d\lambda} = -u_{z_0}\,\ell_{z_0}[\pi_{\lambda}].
  \label{SMeq:pi_lambda_derivative_2}
\end{equation}
Comparing \cref{SMeq:pi_lambda_derivative_2} with \cref{SMeq:vlambda_equation}, and using the zero-mass condition implied by normalization, we obtain
\begin{equation}
  \frac{d\pi_{\lambda}(x)}{d\lambda} = \ell_{z_0}[\pi_{\lambda}]\,v_{\lambda}(x).
  \label{SMeq:pi_lambda_response}
\end{equation}

We next derive the $\lambda$ dependence of $v_{\lambda}(x)$. Differentiating \cref{SMeq:vlambda_equation} with respect to $\lambda$ gives
\begin{equation}
  \frac{d\mathcal{L}_{\lambda}}{d\lambda}v_{\lambda} + \mathcal{L}_{\lambda}\frac{dv_{\lambda}}{d\lambda} = 0.
  \label{SMeq:vlambda_derivative_1}
\end{equation}
Using \cref{SMeq:generator_lambda_derivative}, we obtain
\begin{equation}
  \mathcal{L}_{\lambda}\frac{dv_{\lambda}}{d\lambda} = -u_{z_0}\,\ell_{z_0}[v_{\lambda}].
  \label{SMeq:vlambda_derivative_2}
\end{equation}
Comparing \cref{SMeq:vlambda_derivative_2} with \cref{SMeq:vlambda_equation}, and again using the zero-mass condition, we find
\begin{equation}
  \frac{dv_{\lambda}(x)}{d\lambda} = \ell_{z_0}[v_{\lambda}]\,v_{\lambda}(x).
  \label{SMeq:vlambda_scalar_equation}
\end{equation}
Equation \cref{SMeq:vlambda_scalar_equation} shows that the local response mode remains parallel to itself along the perturbation path. Therefore, for any reference value $\lambda_{\ast}$, there exists a scalar function $M(\lambda)$ such that
\begin{equation}
  v_{\lambda}(x) = M(\lambda)\,v_{\lambda_{\ast}}(x).
  \label{SMeq:vlambda_collinear}
\end{equation}
Substituting \cref{SMeq:vlambda_collinear} into \cref{SMeq:pi_lambda_response}, we obtain
\begin{equation}
  \frac{d\pi_{\lambda}(x)}{d\lambda} = \ell_{z_0}[\pi_{\lambda}]\,M(\lambda)\,v_{\lambda_{\ast}}(x).
  \label{SMeq:pi_lambda_response_collinear}
\end{equation}
Therefore, the full stationary density lies on a one-dimensional affine manifold,
\begin{equation}
  \pi_{\lambda}(x) = \pi_{\lambda_{\ast}}(x) + s(\lambda)\,v_{\lambda_{\ast}}(x),
  \label{SMeq:pi_lambda_affine_manifold}
\end{equation}
where the scalar amplitude $s(\lambda)$ is determined by
\begin{equation}
  \frac{ds(\lambda)}{d\lambda} = \ell_{z_0}[\pi_{\lambda}]\,M(\lambda),
  \qquad
  s(\lambda_{\ast}) = 0.
  \label{SMeq:s_lambda_equation}
\end{equation}
Evaluating \cref{SMeq:pi_lambda_affine_manifold} at two positions $x_1$ and $x_2$ immediately gives
\begin{equation}
  \pi_{\lambda}(x_1) = \chi_{x_1x_2}\,\pi_{\lambda}(x_2) + \gamma_{x_1x_2},
  \label{SMeq:density_mutual_linearity}
\end{equation}
where
\begin{equation}
  \chi_{x_1x_2} \equiv \frac{v_{\lambda_{\ast}}(x_1)}{v_{\lambda_{\ast}}(x_2)} = \frac{\partial_z Z_{\lambda_{\ast}}(x_1|z)\big|_{z=z_0}}{\partial_z Z_{\lambda_{\ast}}(x_2|z)\big|_{z=z_0}},
  \label{SMeq:density_slope_definition}
\end{equation}
and
\begin{equation}
  \gamma_{x_1x_2} \equiv \pi_{\lambda_{\ast}}(x_1)-\chi_{x_1x_2}\pi_{\lambda_{\ast}}(x_2).
  \label{SMeq:density_intercept_definition}
\end{equation}
Both $\chi_{x_1x_2}$ and $\gamma_{x_1x_2}$ are independent of $\lambda$.

\subsection{Mutual linearity of state-current observables}
\label{SMsec:observable_mutual_linearity}

We now extend the above result from the stationary density to general state-current observables. Consider
\begin{equation}
  Q_m(t) = \int dx\,a_m(x)p(x,t) + \int dx\,b_m(x)j(x,t),
  \qquad
  m=1,2.
  \label{SMeq:Qm_definition}
\end{equation}
Its steady-state value under the generator $\mathcal{L}_{\lambda}$ is
\begin{equation}
  Q_{m,\mathrm{ss},\lambda} = \int dx\,a_m(x)\pi_{\lambda}(x) + \int dx\,b_m(x)j_{\mathrm{ss},\lambda}(x).
  \label{SMeq:Qm_stationary_average}
\end{equation}
We now impose the excluded-point condition at the perturbed position $z_0$,
\begin{equation}
  b_m(z_0)=0,
  \qquad
  m=1,2.
  \label{SMeq:excluded_point_condition_global}
\end{equation}
Using \cref{SMeq:stationary_current_lambda}, together with the definition of $\mathcal{J}_{\lambda}$ in \cref{SMeq:Jlambda_Llambda}, we can write
\begin{equation}
  j_{\mathrm{ss},\lambda}(x) = \mathcal{J}_{0}\pi_{\lambda}(x) + \lambda\delta(x-z_0)\mathcal{K}_{\phi}\pi_{\lambda}(x).
  \label{SMeq:current_lambda_split}
\end{equation}
Substituting \cref{SMeq:current_lambda_split} into \cref{SMeq:Qm_stationary_average}, we obtain
\begin{equation}
  Q_{m,\mathrm{ss},\lambda} = \int dx\,a_m(x)\pi_{\lambda}(x) + \int dx\,b_m(x)\mathcal{J}_{0}\pi_{\lambda}(x) + \lambda b_m(z_0)\bigl(\mathcal{K}_{\phi}\pi_{\lambda}\bigr)(z_0).
  \label{SMeq:Qm_stationary_average_split}
\end{equation}
By the excluded-point condition \cref{SMeq:excluded_point_condition_global}, the last term vanishes identically, and therefore
\begin{equation}
  Q_{m,\mathrm{ss},\lambda} = \int dx\,a_m(x)\pi_{\lambda}(x) + \int dx\,b_m(x)\mathcal{J}_{0}\pi_{\lambda}(x).
  \label{SMeq:Qm_stationary_average_excluded}
\end{equation}
Under boundary conditions for which the integration-by-parts boundary term vanishes, one has
\begin{equation}
  \int dx\,b_m(x)\mathcal{J}_{0}f(x) = \int dx\,\bigl[\mathcal{J}_{0}^{\dagger}b_m(x)\bigr]f(x).
  \label{SMeq:Qm_adjoint_identity}
\end{equation}
Hence \cref{SMeq:Qm_stationary_average_excluded} becomes
\begin{equation}
  Q_{m,\mathrm{ss},\lambda} = \int dx\,c_m(x)\pi_{\lambda}(x),
  \qquad
  c_m(x) \equiv a_m(x)+\mathcal{J}_{0}^{\dagger}b_m(x).
  \label{SMeq:Qm_density_functional}
\end{equation}
Substituting the affine density manifold in \cref{SMeq:pi_lambda_affine_manifold} into \cref{SMeq:Qm_density_functional}, we obtain
\begin{equation}
  Q_{m,\mathrm{ss},\lambda} = A_m + s(\lambda)B_m,
  \label{SMeq:Qm_affine_in_s}
\end{equation}
where
\begin{equation}
  A_m \equiv \int dx\,c_m(x)\pi_{\lambda_{\ast}}(x),
  \qquad
  B_m \equiv \int dx\,c_m(x)v_{\lambda_{\ast}}(x).
  \label{SMeq:Qm_affine_coefficients}
\end{equation}
Therefore, for any two such observables with $B_2\neq 0$, eliminating the common scalar $s(\lambda)$ gives
\begin{equation}
  Q_{1,\mathrm{ss},\lambda} = \chi\,Q_{2,\mathrm{ss},\lambda} + \gamma,
  \label{SMeq:observable_mutual_linearity}
\end{equation}
with
\begin{equation}
  \chi \equiv \frac{B_1}{B_2},
  \qquad
  \gamma \equiv A_1 - \frac{B_1}{B_2}A_2.
  \label{SMeq:observable_mutual_linearity_coefficients}
\end{equation}
Equation \cref{SMeq:observable_mutual_linearity} is the exact steady-state mutual linearity for excluded-point state-current observables in overdamped Langevin dynamics.

\section{Laplace-domain Mutual Linearity}
\label{SMsec:laplace}

We now show that the same one-dimensional response structure persists after the Laplace transformation. This provides the continuous analogue of the frequency-domain mutual linearity. The basic idea is parallel to the steady-state derivation in \cref{SMsec:distribution_mutual_linearity,SMsec:observable_mutual_linearity}: once the perturbation is localized at a fixed point, the $\lambda$ dependence of the dynamics remains rank-one, and the Laplace-transformed dynamics is again confined to a one-dimensional affine manifold.

We consider the same one-parameter family of localized perturbations introduced in \cref{SMeq:philambda,SMeq:Jlambda_Llambda}. Let the initial density $p_0(x)$ be independent of $\lambda$, and let $p_{\lambda}(x,t)$ evolve according to
\begin{equation}
  \partial_t p_{\lambda}(x,t) = \mathcal{L}_{\lambda}p_{\lambda}(x,t),
  \qquad
  p_{\lambda}(x,0) = p_0(x).
  \label{SMeq:laplace_time_evolution}
\end{equation}
For $\operatorname{Re}\omega > 0$, we define the Laplace transform of the time-dependent density by
\begin{equation}
  \hat{p}_{\lambda}(x,\omega) \equiv \int_0^{\infty} e^{-\omega t}p_{\lambda}(x,t)\,dt.
  \label{SMeq:p_hat_definition}
\end{equation}
Applying the Laplace transform to \cref{SMeq:laplace_time_evolution}, we obtain
\begin{equation}
  (\omega - \mathcal{L}_{\lambda})\hat{p}_{\lambda}(x,\omega) = p_0(x).
  \label{SMeq:p_hat_resolvent_equation}
\end{equation}

To analyze the $\lambda$ dependence of $\hat{p}_{\lambda}(x,\omega)$, we introduce the Laplace-domain propagator kernel
\begin{equation}
  \hat{Z}_{\lambda}(x|y;\omega) \equiv \int_0^{\infty} e^{-\omega t}P_{\lambda}(x,t|y,0)\,dt,
  \label{SMeq:Z_hat_definition}
\end{equation}
where $P_{\lambda}(x,t|y,0)$ is the propagator generated by $\mathcal{L}_{\lambda}$. By construction, it satisfies
\begin{equation}
  (\omega - \mathcal{L}_{\lambda,x})\hat{Z}_{\lambda}(x|y;\omega) = \delta(x-y).
  \label{SMeq:Z_hat_equation}
\end{equation}
We then define the Laplace-domain local response mode at the perturbed point $z_0$ by
\begin{equation}
  \hat{v}_{\lambda}(x,\omega) \equiv \partial_z \hat{Z}_{\lambda}(x|z;\omega)\big|_{z=z_0}.
  \label{SMeq:v_hat_definition}
\end{equation}
Differentiating \cref{SMeq:Z_hat_equation} with respect to $z$ and setting $z=z_0$, we obtain
\begin{equation}
  (\omega - \mathcal{L}_{\lambda})\hat{v}_{\lambda}(x,\omega) = u_{z_0}(x),
  \qquad
  u_{z_0}(x) \equiv -\partial_x\delta(x-z_0).
  \label{SMeq:v_hat_equation}
\end{equation}

As in the steady-state case, the derivative of the generator with respect to $\lambda$ is rank-one. For any test function $f(x)$,
\begin{equation}
  \frac{d\mathcal{L}_{\lambda}}{d\lambda}f = u_{z_0}\,\ell_{z_0}[f],
  \qquad
  \ell_{z_0}[f] \equiv \bigl(\mathcal{K}_{\phi}f\bigr)(z_0).
  \label{SMeq:laplace_generator_lambda_derivative}
\end{equation}
We first determine how $\hat{v}_{\lambda}(x,\omega)$ changes with $\lambda$. Differentiating \cref{SMeq:v_hat_equation} with respect to $\lambda$, we find
\begin{equation}
  -\frac{d\mathcal{L}_{\lambda}}{d\lambda}\hat{v}_{\lambda} + (\omega-\mathcal{L}_{\lambda})\frac{d\hat{v}_{\lambda}}{d\lambda} = 0.
  \label{SMeq:v_hat_derivative_1}
\end{equation}
Using \cref{SMeq:laplace_generator_lambda_derivative}, this becomes
\begin{equation}
  (\omega-\mathcal{L}_{\lambda})\frac{d\hat{v}_{\lambda}}{d\lambda} = u_{z_0}\,\ell_{z_0}[\hat{v}_{\lambda}].
  \label{SMeq:v_hat_derivative_2}
\end{equation}
Comparing \cref{SMeq:v_hat_derivative_2} with \cref{SMeq:v_hat_equation}, we obtain
\begin{equation}
  \frac{d\hat{v}_{\lambda}(x,\omega)}{d\lambda} = \ell_{z_0}[\hat{v}_{\lambda}]\,\hat{v}_{\lambda}(x,\omega).
  \label{SMeq:v_hat_scalar_equation}
\end{equation}
Thus, for any reference point $\lambda_{\ast}$, there exists a scalar function $\hat{M}(\lambda,\omega)$ such that
\begin{equation}
  \hat{v}_{\lambda}(x,\omega) = \hat{M}(\lambda,\omega)\,\hat{v}_{\lambda_{\ast}}(x,\omega).
  \label{SMeq:v_hat_proportionality}
\end{equation}

We now derive the $\lambda$ dependence of the Laplace-transformed density itself. Differentiating \cref{SMeq:p_hat_resolvent_equation} with respect to $\lambda$ yields
\begin{equation}
  -\frac{d\mathcal{L}_{\lambda}}{d\lambda}\hat{p}_{\lambda} + (\omega-\mathcal{L}_{\lambda})\frac{d\hat{p}_{\lambda}}{d\lambda} = 0.
  \label{SMeq:p_hat_derivative_1}
\end{equation}
Using \cref{SMeq:laplace_generator_lambda_derivative}, we obtain
\begin{equation}
  (\omega-\mathcal{L}_{\lambda})\frac{d\hat{p}_{\lambda}}{d\lambda} = u_{z_0}\,\ell_{z_0}[\hat{p}_{\lambda}].
  \label{SMeq:p_hat_derivative_2}
\end{equation}
Comparing \cref{SMeq:p_hat_derivative_2} with \cref{SMeq:v_hat_equation}, we find
\begin{equation}
  \frac{d\hat{p}_{\lambda}(x,\omega)}{d\lambda} = \ell_{z_0}[\hat{p}_{\lambda}]\,\hat{v}_{\lambda}(x,\omega).
  \label{SMeq:p_hat_response}
\end{equation}
Substituting \cref{SMeq:v_hat_proportionality} into \cref{SMeq:p_hat_response}, we conclude that the Laplace-transformed density also evolves along a one-dimensional affine manifold:
\begin{equation}
  \hat{p}_{\lambda}(x,\omega) = \hat{p}_{\lambda_{\ast}}(x,\omega) + \hat{s}(\lambda,\omega)\,\hat{v}_{\lambda_{\ast}}(x,\omega),
  \label{SMeq:p_hat_affine}
\end{equation}
where $\hat{s}(\lambda,\omega)$ is a scalar function determined by $d\hat{s}(\lambda,\omega)/d\lambda = \ell_{z_0}[\hat{p}_{\lambda}]\,\hat{M}(\lambda,\omega)$.

We next turn to the Laplace transform of observables. For each $Q_m$, we define
\begin{equation}
  Q_{m,\lambda}(t) \equiv \int dx\,a_m(x)p_{\lambda}(x,t) + \int dx\,b_m(x)j_{\lambda}(x,t),
  \label{SMeq:Qm_time_average}
\end{equation}
where
\begin{equation}
  j_{\lambda}(x,t) \equiv \mathcal{J}_{\lambda}p_{\lambda}(x,t).
  \label{SMeq:current_time_dependent}
\end{equation}
We then define the Laplace transform of the time-dependent observable by
\begin{equation}
  \hat{Q}_m(\omega)\big|_{\lambda} \equiv \int_0^{\infty} e^{-\omega t}Q_{m,\lambda}(t)\,dt.
  \label{SMeq:Qm_hat_definition}
\end{equation}

Assume again that the observables exclude the perturbed point,
\begin{equation}
  b_m(z_0)=0,
  \qquad
  m=1,2.
  \label{SMeq:Qm_hat_excluded_point}
\end{equation}
Then, using $\mathcal{J}_{\lambda}=\mathcal{J}_{0}+\lambda\delta(x-z_0)\mathcal{K}_{\phi}$ together with \cref{SMeq:Qm_hat_excluded_point}, the contact contribution vanishes exactly, and the same adjoint manipulation as in \cref{SMsec:observable_mutual_linearity} gives
\begin{equation}
  \hat{Q}_m(\omega)\big|_{\lambda} = \int dx\,c_m(x)\hat{p}_{\lambda}(x,\omega),
  \qquad
  c_m(x) \equiv a_m(x)+\mathcal{J}_{0}^{\dagger}b_m(x).
  \label{SMeq:Qm_hat_density_functional}
\end{equation}
Substituting \cref{SMeq:p_hat_affine} into \cref{SMeq:Qm_hat_density_functional}, we obtain
\begin{equation}
  \hat{Q}_m(\omega)\big|_{\lambda} = \hat{A}_m(\omega) + \hat{s}(\lambda,\omega)\hat{B}_m(\omega),
  \label{SMeq:Qm_hat_affine}
\end{equation}
where
\begin{equation}
  \hat{A}_m(\omega) \equiv \int dx\,c_m(x)\hat{p}_{\lambda_{\ast}}(x,\omega),
  \qquad
  \hat{B}_m(\omega) \equiv \int dx\,c_m(x)\hat{v}_{\lambda_{\ast}}(x,\omega).
  \label{SMeq:Qm_hat_affine_coefficients}
\end{equation}
Therefore, if $\hat{B}_2(\omega)\neq 0$, eliminating the common scalar $\hat{s}(\lambda,\omega)$ gives
\begin{equation}
  \hat{Q}_1(\omega)\big|_{\lambda} = \hat{\chi}(\omega)\,\hat{Q}_2(\omega)\big|_{\lambda} + \hat{\gamma}(\omega),
  \label{SMeq:laplace_mutual_linearity}
\end{equation}
with
\begin{equation}
  \hat{\chi}(\omega) \equiv \frac{\hat{B}_1(\omega)}{\hat{B}_2(\omega)},
  \qquad
  \hat{\gamma}(\omega) \equiv \hat{A}_1(\omega) - \frac{\hat{B}_1(\omega)}{\hat{B}_2(\omega)}\hat{A}_2(\omega).
  \label{SMeq:laplace_mutual_linearity_coefficients}
\end{equation}
Finally, the steady-state result is recovered from the small-$\omega$ limit. Since
\begin{equation}
  \hat{p}_{\lambda}(x,\omega) = \frac{\pi_{\lambda}(x)}{\omega} + O(1)
  \qquad
  (\omega \to 0^{+}),
  \label{SMeq:small_omega_limit}
\end{equation}
the affine relation in \cref{SMeq:laplace_mutual_linearity} reduces to the steady-state affine relation in \cref{SMeq:observable_mutual_linearity}.

\section{Multidimensional Mutual Linearity}
\label{SMsec:multi}

In this section, we extend the same rank-one response structure to multidimensional overdamped Langevin dynamics. The key point is that, for a perturbation of a single local channel, the derivative of the generator with respect to the perturbation strength remains rank-one. As a result, the two structural relations
\begin{equation}
  \frac{d\pi_{\lambda}}{d\lambda} \propto v_{\lambda},
  \qquad
  \frac{dv_{\lambda}}{d\lambda} \propto v_{\lambda},
  \label{SMeq:multi_core_structure}
\end{equation}
continue to hold in an arbitrary dimension.

We consider a $d$-dimensional overdamped Langevin system described by
\begin{equation}
  \dot{\bm{x}}_t = \bm{M}(\bm{x}_t)\bm{F}(\bm{x}_t) + \sqrt{2}\,\bm{B}(\bm{x}_t)\,\star\,\bm{\xi}_t,
  \qquad
  \bm{D}(\bm{x}) \equiv \bm{B}(\bm{x})\bm{B}(\bm{x})^{\mathsf T} = \bm{M}(\bm{x})\bm{T}(\bm{x}),
  \label{SMeq:multi_langevin}
\end{equation}
where $\bm{x}_t=(x_1(t),\ldots,x_d(t))^{\mathsf T}$ is the stochastic state variable, $\bm{F}(\bm{x})$ is the systematic force, $\bm{M}(\bm{x})$ is the mobility matrix, and $\bm{T}(\bm{x})=\diag\bigl(T_1(\bm{x}),\ldots,T_d(\bm{x})\bigr)$ is the temperature matrix. The noise $\bm{\xi}_t$ is a $d$-component Gaussian white noise with zero mean and covariance $\langle \xi_i(t)\xi_j(t')\rangle = \delta_{ij}\delta(t-t')$. The corresponding probability-current operator and Fokker-Planck generator are
\begin{equation}
  \bm{\mathcal{J}}f = \bm{M}(\bm{x})\bigl[\bm{F}(\bm{x})f-\bm{T}(\bm{x})\nabla_{\bm{x}}f\bigr],
  \qquad
  \mathcal{L}f = -\nabla_{\bm{x}}\cdot \bm{\mathcal{J}}f.
  \label{SMeq:multi_current_generator}
\end{equation}
We assume that the dynamics admits a unique stationary density $\pi(\bm{x})$ satisfying $\mathcal{L}\pi(\bm{x})=0$, and denote the corresponding stationary current by $\bm{j}_{\mathrm{ss}}(\bm{x})=\bm{\mathcal{J}}\pi(\bm{x})$.

\subsection{Response to local perturbations}
\label{SMsec:multi_response}

We perturb a single scalar channel $\phi \in \{F_i, M_{ij}, T_i\}$ at a fixed position $\bm{z}$. Let $\bm{e}_i$ denote the unit vector with components $[\bm{e}_i]_m = \delta_{im}$. The perturbed current operator can be written as
\begin{equation}
  \bm{\mathcal{J}}_{\phi}^{\epsilon,\bm{z}} \equiv \bm{\mathcal{J}} + \epsilon\,\delta(\bm{x}-\bm{z})\,\bm{\mathcal{K}}_{\phi},
  \label{SMeq:multi_perturbed_current}
\end{equation}
where the local channel can be factorized as
\begin{equation}
  \bm{\mathcal{K}}_{\phi} = \bm{a}_{\phi}(\bm{x})\,\mathcal{K}_{\phi}.
  \label{SMeq:multi_channel_factorization}
\end{equation}
For the three single-channel perturbations, one has
\begin{subequations}
\begin{align}
  \bm{a}_{F_i}(\bm{x}) &= \bm{M}(\bm{x})\bm{e}_i,
  &
  \mathcal{K}_{F_i} &= 1,
  \label{SMeq:multi_channel_force}
  \\
  \bm{a}_{M_{ij}}(\bm{x}) &= \bm{e}_i,
  &
  \mathcal{K}_{M_{ij}} &= F_j(\bm{x}) - T_j(\bm{x})\partial_{x_j},
  \label{SMeq:multi_channel_mobility}
  \\
  \bm{a}_{T_i}(\bm{x}) &= -\bm{M}(\bm{x})\bm{e}_i,
  &
  \mathcal{K}_{T_i} &= \partial_{x_i}.
  \label{SMeq:multi_channel_temperature}
\end{align}
\end{subequations}
Thus the perturbation acts through a fixed vector direction $\bm{a}_{\phi}$ and a scalar local operator $\mathcal{K}_{\phi}$.

Let $P(\bm{x},t|\bm{z},0)$ be the propagator generated by $\mathcal{L}$, and define the multidimensional fundamental kernel
\begin{equation}
  Z(\bm{x}|\bm{z}) \equiv \int_0^{\infty} \bigl[P(\bm{x},t|\bm{z},0)-\pi(\bm{x})\bigr]dt.
  \label{SMeq:multi_fundamental_kernel}
\end{equation}
It satisfies
\begin{equation}
  \mathcal{L}_{\bm{x}}Z(\bm{x}|\bm{z}) = \pi(\bm{x}) - \delta(\bm{x}-\bm{z}),
  \qquad
  \int d\bm{x}\,Z(\bm{x}|\bm{z})=0.
  \label{SMeq:multi_fundamental_kernel_equation}
\end{equation}
For the perturbed channel $\phi$, we define the scalar local response mode
\begin{equation}
  v_{\phi,\bm{z}}(\bm{x}) \equiv \bm{a}_{\phi}(\bm{z})\cdot \nabla_{\bm{z}} Z(\bm{x}|\bm{z}).
  \label{SMeq:multi_local_mode}
\end{equation}
Differentiating \cref{SMeq:multi_fundamental_kernel_equation} along the direction $\bm{a}_{\phi}(\bm{z})$ gives
\begin{equation}
  \mathcal{L}v_{\phi,\bm{z}}(\bm{x}) = -u_{\phi,\bm{z}}(\bm{x}),
  \qquad
  u_{\phi,\bm{z}}(\bm{x}) \equiv -\bm{a}_{\phi}(\bm{z})\cdot \nabla_{\bm{x}}\delta(\bm{x}-\bm{z}),
  \label{SMeq:multi_local_mode_equation}
\end{equation}
together with
\begin{equation}
  \int d\bm{x}\,v_{\phi,\bm{z}}(\bm{x}) = 0.
  \label{SMeq:multi_local_mode_zero_mass}
\end{equation}

We now derive the response to an infinitesimal local perturbation. Expanding the stationary density as
\begin{equation}
  \pi_{\phi}^{\epsilon,\bm{z}}(\bm{x}) = \pi(\bm{x}) + \epsilon\,q_{\phi}(\bm{x};\bm{z}) + O(\epsilon^2),
  \label{SMeq:multi_pi_expansion}
\end{equation}
and inserting this into the stationary equation generated by \cref{SMeq:multi_perturbed_current}, we obtain
\begin{equation}
  \mathcal{L}q_{\phi}(\bm{x};\bm{z}) = -u_{\phi,\bm{z}}(\bm{x})\,\ell_{\phi,\bm{z}}[\pi],
  \qquad
  \ell_{\phi,\bm{z}}[f] \equiv \bigl(\mathcal{K}_{\phi}f\bigr)(\bm{z}).
  \label{SMeq:multi_q_equation}
\end{equation}
Comparing \cref{SMeq:multi_q_equation} with \cref{SMeq:multi_local_mode_equation}, we immediately obtain
\begin{equation}
  \frac{\delta \pi(\bm{x})}{\delta \phi(\bm{z})} = N_{\phi}(\bm{z})\,v_{\phi,\bm{z}}(\bm{x}),
  \qquad
  N_{\phi}(\bm{z}) \equiv \ell_{\phi,\bm{z}}[\pi].
  \label{SMeq:multi_density_response}
\end{equation}
Similarly, the stationary current response takes the form
\begin{equation}
  \frac{\delta \bm{j}_{\mathrm{ss}}(\bm{x})}{\delta \phi(\bm{z})} = N_{\phi}(\bm{z})\Bigl[\bm{a}_{\phi}(\bm{z})\delta(\bm{x}-\bm{z}) + \bm{\mathcal{J}}v_{\phi,\bm{z}}(\bm{x})\Bigr].
  \label{SMeq:multi_current_response}
\end{equation}
Thus the same one-kernel structure survives in arbitrary dimension.

\subsection{Mutual linearity of the stationary density}
\label{SMsec:multi_density}

We now turn to a finite perturbation of the same local channel. Fix a point $\bm{z}_0$ and a single scalar channel $\phi \in \{F_i, M_{ij}, T_i\}$. We consider the one-parameter family
\begin{equation}
  \phi_{\lambda}(\bm{x}) = \phi_{0}(\bm{x}) + \lambda\,\delta(\bm{x}-\bm{z}_0).
  \label{SMeq:multi_philambda}
\end{equation}
The corresponding current operator and generator are
\begin{equation}
  \bm{\mathcal{J}}_{\lambda} = \bm{\mathcal{J}}_{0} + \lambda\,\delta(\bm{x}-\bm{z}_0)\bm{a}_{\phi}(\bm{x})\mathcal{K}_{\phi},
  \qquad
  \mathcal{L}_{\lambda} = -\nabla_{\bm{x}}\cdot \bm{\mathcal{J}}_{\lambda}.
  \label{SMeq:multi_Jlambda_Llambda}
\end{equation}
For each $\lambda$, let $\pi_{\lambda}(\bm{x})$ be the unique stationary density satisfying
\begin{equation}
  \mathcal{L}_{\lambda}\pi_{\lambda}(\bm{x}) = 0,
  \qquad
  \int d\bm{x}\,\pi_{\lambda}(\bm{x})=1.
  \label{SMeq:multi_stationary_density}
\end{equation}

Let $Z_{\lambda}(\bm{x}|\bm{z})$ be the corresponding fundamental kernel,
\begin{equation}
  Z_{\lambda}(\bm{x}|\bm{z}) \equiv \int_0^{\infty} \bigl[P_{\lambda}(\bm{x},t|\bm{z},0)-\pi_{\lambda}(\bm{x})\bigr]dt,
  \label{SMeq:multi_fundamental_kernel_lambda}
\end{equation}
which satisfies
\begin{equation}
  \mathcal{L}_{\lambda,\bm{x}} Z_{\lambda}(\bm{x}|\bm{z}) = \pi_{\lambda}(\bm{x}) - \delta(\bm{x}-\bm{z}),
  \qquad
  \int d\bm{x}\,Z_{\lambda}(\bm{x}|\bm{z})=0.
  \label{SMeq:multi_fundamental_kernel_lambda_equation}
\end{equation}
We define the local response mode associated with the perturbed channel by
\begin{equation}
  v_{\lambda,\phi}(\bm{x}) \equiv \bm{a}_{\phi}(\bm{z}_0)\cdot \nabla_{\bm{z}} Z_{\lambda}(\bm{x}|\bm{z})\big|_{\bm{z}=\bm{z}_0}.
  \label{SMeq:multi_vlambda_definition}
\end{equation}
Differentiating \cref{SMeq:multi_fundamental_kernel_lambda_equation} with respect to $\bm{z}$ along the direction $\bm{a}_{\phi}(\bm{z}_0)$, we obtain
\begin{equation}
  \mathcal{L}_{\lambda}v_{\lambda,\phi}(\bm{x}) = -u_{\phi,\bm{z}_0}(\bm{x}),
  \qquad
  u_{\phi,\bm{z}_0}(\bm{x}) \equiv -\bm{a}_{\phi}(\bm{z}_0)\cdot \nabla_{\bm{x}}\delta(\bm{x}-\bm{z}_0),
  \label{SMeq:multi_vlambda_equation}
\end{equation}
with
\begin{equation}
  \int d\bm{x}\,v_{\lambda,\phi}(\bm{x}) = 0.
  \label{SMeq:multi_vlambda_zero_mass}
\end{equation}

For any test function $f(\bm{x})$, the derivative of the generator with respect to $\lambda$ is
\begin{equation}
  \frac{d\mathcal{L}_{\lambda}}{d\lambda}f = -\nabla_{\bm{x}}\cdot\Bigl[\delta(\bm{x}-\bm{z}_0)\bm{a}_{\phi}(\bm{z}_0)\Bigr] \, \ell_{\phi,\bm{z}_0}[f]
  = u_{\phi,\bm{z}_0}\,\ell_{\phi,\bm{z}_0}[f],
  \label{SMeq:multi_generator_derivative}
\end{equation}
where
\begin{equation}
  \ell_{\phi,\bm{z}_0}[f] \equiv \bigl(\mathcal{K}_{\phi}f\bigr)(\bm{z}_0).
  \label{SMeq:multi_ell_definition}
\end{equation}
Differentiating \cref{SMeq:multi_stationary_density} with respect to $\lambda$ and using \cref{SMeq:multi_generator_derivative}, we obtain
\begin{equation}
  \mathcal{L}_{\lambda}\frac{d\pi_{\lambda}}{d\lambda} = -u_{\phi,\bm{z}_0}\,\ell_{\phi,\bm{z}_0}[\pi_{\lambda}].
  \label{SMeq:multi_pi_derivative}
\end{equation}
Comparing \cref{SMeq:multi_pi_derivative} with \cref{SMeq:multi_vlambda_equation} and using the zero-mass condition implied by normalization, we find
\begin{equation}
  \frac{d\pi_{\lambda}(\bm{x})}{d\lambda} = \ell_{\phi,\bm{z}_0}[\pi_{\lambda}]\,v_{\lambda,\phi}(\bm{x}).
  \label{SMeq:multi_pi_response}
\end{equation}
Likewise, differentiating \cref{SMeq:multi_vlambda_equation} with respect to $\lambda$ gives
\begin{equation}
  \mathcal{L}_{\lambda}\frac{dv_{\lambda,\phi}}{d\lambda} = -u_{\phi,\bm{z}_0}\,\ell_{\phi,\bm{z}_0}[v_{\lambda,\phi}],
  \label{SMeq:multi_v_derivative}
\end{equation}
so that
\begin{equation}
  \frac{dv_{\lambda,\phi}(\bm{x})}{d\lambda} = \ell_{\phi,\bm{z}_0}[v_{\lambda,\phi}]\,v_{\lambda,\phi}(\bm{x}).
  \label{SMeq:multi_v_response}
\end{equation}
Thus the local response mode remains parallel to itself along the perturbation path. For any reference value $\lambda_{\ast}$, there exists a scalar function $M(\lambda)$ such that
\begin{equation}
  v_{\lambda,\phi}(\bm{x}) = M(\lambda)\,v_{\lambda_{\ast},\phi}(\bm{x}).
  \label{SMeq:multi_v_collinear}
\end{equation}
Substituting \cref{SMeq:multi_v_collinear} into \cref{SMeq:multi_pi_response}, we conclude that the full stationary density lies on a one-dimensional affine manifold,
\begin{equation}
  \pi_{\lambda}(\bm{x}) = \pi_{\lambda_{\ast}}(\bm{x}) + s(\lambda)\,v_{\lambda_{\ast},\phi}(\bm{x}),
  \label{SMeq:multi_pi_affine_manifold}
\end{equation}
with
\begin{equation}
  \frac{ds(\lambda)}{d\lambda} = \ell_{\phi,\bm{z}_0}[\pi_{\lambda}]\,M(\lambda),
  \qquad
  s(\lambda_{\ast}) = 0.
  \label{SMeq:multi_s_equation}
\end{equation}
Therefore, for any two observation points $\bm{x}_1$ and $\bm{x}_2$ such that $v_{\lambda_{\ast},\phi}(\bm{x}_2)\neq 0$, one has
\begin{equation}
  \pi_{\lambda}(\bm{x}_1) = \chi_{\bm{x}_1\bm{x}_2}\,\pi_{\lambda}(\bm{x}_2) + \gamma_{\bm{x}_1\bm{x}_2},
  \label{SMeq:multi_density_mutual_linearity}
\end{equation}
where
\begin{equation}
  \chi_{\bm{x}_1\bm{x}_2} \equiv \frac{v_{\lambda_{\ast},\phi}(\bm{x}_1)}{v_{\lambda_{\ast},\phi}(\bm{x}_2)},
  \qquad
  \gamma_{\bm{x}_1\bm{x}_2} \equiv \pi_{\lambda_{\ast}}(\bm{x}_1)-\chi_{\bm{x}_1\bm{x}_2}\pi_{\lambda_{\ast}}(\bm{x}_2).
  \label{SMeq:multi_density_mutual_linearity_coefficients}
\end{equation}
Thus, for a perturbation of a single local channel, the multidimensional stationary density remains exactly one-dimensional in the sense of mutual linearity. The dimension of the state space does not enlarge the response manifold.

\subsection{Mutual linearity of state-current observables}
\label{SMsec:multi_observable}

We now consider multidimensional state-current observables of the form
\begin{equation}
  Q(t) = \int d\bm{x}\,a(\bm{x})p(\bm{x},t) + \int d\bm{x}\,\bm{b}(\bm{x})\cdot \bm{j}(\bm{x},t),
  \label{SMeq:multi_observable}
\end{equation}
where $a(\bm{x})$ is a scalar weight and $\bm{b}(\bm{x})$ is a vector weight. In the steady state,
\begin{equation}
  Q_{\mathrm{ss},\lambda} = \int d\bm{x}\,a(\bm{x})\pi_{\lambda}(\bm{x}) + \int d\bm{x}\,\bm{b}(\bm{x})\cdot \bm{j}_{\mathrm{ss},\lambda}(\bm{x}).
  \label{SMeq:multi_observable_ss}
\end{equation}
For the single-channel perturbation considered above, the direct local current contribution vanishes provided that
\begin{equation}
  \bm{b}(\bm{z}_0)\cdot \bm{a}_{\phi}(\bm{z}_0)=0.
  \label{SMeq:multi_excluded_point_condition}
\end{equation}
A stronger but simpler sufficient condition is $\bm{b}(\bm{z}_0)=\bm{0}$.

Using \cref{SMeq:multi_Jlambda_Llambda}, we can write
\begin{equation}
  \bm{j}_{\mathrm{ss},\lambda}(\bm{x}) = \bm{\mathcal{J}}_{0}\pi_{\lambda}(\bm{x}) + \lambda\delta(\bm{x}-\bm{z}_0)\bm{a}_{\phi}(\bm{z}_0)\,\ell_{\phi,\bm{z}_0}[\pi_{\lambda}].
  \label{SMeq:multi_current_split}
\end{equation}
Substituting \cref{SMeq:multi_current_split} into \cref{SMeq:multi_observable_ss} and using \cref{SMeq:multi_excluded_point_condition}, we obtain
\begin{equation}
  Q_{\mathrm{ss},\lambda} = \int d\bm{x}\,a(\bm{x})\pi_{\lambda}(\bm{x}) + \int d\bm{x}\,\bm{b}(\bm{x})\cdot \bm{\mathcal{J}}_{0}\pi_{\lambda}(\bm{x}).
  \label{SMeq:multi_observable_ss_excluded}
\end{equation}
Under boundary conditions for which the boundary term vanishes, we define the adjoint through
\begin{equation}
  \int d\bm{x}\,\bm{b}(\bm{x})\cdot \bm{\mathcal{J}}_{0}f(\bm{x}) = \int d\bm{x}\,\bigl[\bm{\mathcal{J}}_{0}^{\dagger}\bm{b}(\bm{x})\bigr]f(\bm{x}),
  \label{SMeq:multi_adjoint_identity}
\end{equation}
where
\begin{equation}
  \bm{\mathcal{J}}^{\dagger}\bm{b}(\bm{x}) \equiv \bigl[\bm{M}(\bm{x})\bm{F}(\bm{x})\bigr]\cdot \bm{b}(\bm{x}) + \nabla_{\bm{x}}\cdot \bigl[\bm{D}(\bm{x})\bm{b}(\bm{x})\bigr].
  \label{SMeq:multi_adjoint_definition}
\end{equation}
Hence
\begin{equation}
  Q_{\mathrm{ss},\lambda} = \int d\bm{x}\,c_Q(\bm{x})\pi_{\lambda}(\bm{x}),
  \qquad
  c_Q(\bm{x}) \equiv a(\bm{x}) + \bm{\mathcal{J}}_{0}^{\dagger}\bm{b}(\bm{x}).
  \label{SMeq:multi_observable_density_functional}
\end{equation}
Substituting \cref{SMeq:multi_pi_affine_manifold} into \cref{SMeq:multi_observable_density_functional}, we find
\begin{equation}
  Q_{m,\mathrm{ss},\lambda} = A_m + s(\lambda)B_m,
  \label{SMeq:multi_observable_affine}
\end{equation}
with
\begin{equation}
  A_m \equiv \int d\bm{x}\,c_m(\bm{x})\pi_{\lambda_{\ast}}(\bm{x}),
  \qquad
  B_m \equiv \int d\bm{x}\,c_m(\bm{x})v_{\lambda_{\ast},\phi}(\bm{x}).
  \label{SMeq:multi_observable_affine_coefficients}
\end{equation}
Therefore, for any two excluded-point multidimensional state-current observables with $B_2\neq 0$, eliminating $s(\lambda)$ yields
\begin{equation}
  Q_{1,\mathrm{ss},\lambda} = \chi\,Q_{2,\mathrm{ss},\lambda} + \gamma,
  \label{SMeq:multi_observable_mutual_linearity}
\end{equation}
where
\begin{equation}
  \chi \equiv \frac{B_1}{B_2},
  \qquad
  \gamma \equiv A_1 - \frac{B_1}{B_2}A_2.
  \label{SMeq:multi_observable_mutual_linearity_coefficients}
\end{equation}
Thus the exact steady-state mutual linearity extends directly to multidimensional overdamped Langevin dynamics, provided that the perturbation acts through a single local channel. If several independent local channels are perturbed simultaneously, the response manifold need not remain one-dimensional.

\subsection{Laplace-domain mutual linearity}
\label{SMsec:multi_laplace}

We now extend the Laplace-domain mutual linearity to multidimensional overdamped Langevin dynamics. We keep the same perturbation channel $\phi \in \{F_i, M_{ij}, T_i\}$ and the same perturbation point $\bm{z}_0$, and let $p_{\lambda}(\bm{x},t)$ evolve under the time-independent generator $\mathcal{L}_{\lambda}$ from an initial density $p_0(\bm{x})$ that is independent of $\lambda$:
\begin{equation}
  \partial_t p_{\lambda}(\bm{x},t) = \mathcal{L}_{\lambda}p_{\lambda}(\bm{x},t),
  \qquad
  p_{\lambda}(\bm{x},0) = p_0(\bm{x}).
  \label{SMeq:multi_laplace_time_evolution}
\end{equation}
For $\operatorname{Re}\omega > 0$, we define the Laplace transform of the time-dependent density by
\begin{equation}
  \hat{p}_{\lambda}(\bm{x},\omega) \equiv \int_0^{\infty} e^{-\omega t}p_{\lambda}(\bm{x},t)\,dt.
  \label{SMeq:multi_p_hat_definition}
\end{equation}
Applying the Laplace transform to \cref{SMeq:multi_laplace_time_evolution}, we obtain
\begin{equation}
  (\omega - \mathcal{L}_{\lambda})\hat{p}_{\lambda}(\bm{x},\omega) = p_0(\bm{x}).
  \label{SMeq:multi_p_hat_resolvent_equation}
\end{equation}

To analyze the $\lambda$ dependence of $\hat{p}_{\lambda}(\bm{x},\omega)$, we introduce the Laplace-domain propagator kernel
\begin{equation}
  \hat{Z}_{\lambda}(\bm{x}|\bm{y};\omega) \equiv \int_0^{\infty} e^{-\omega t}P_{\lambda}(\bm{x},t|\bm{y},0)\,dt,
  \label{SMeq:multi_Z_hat_definition}
\end{equation}
which satisfies
\begin{equation}
  (\omega - \mathcal{L}_{\lambda,\bm{x}})\hat{Z}_{\lambda}(\bm{x}|\bm{y};\omega) = \delta(\bm{x}-\bm{y}).
  \label{SMeq:multi_Z_hat_equation}
\end{equation}
We then define the channel-adapted Laplace-domain local response mode by
\begin{equation}
  \hat{v}_{\lambda,\phi}(\bm{x},\omega) \equiv \bm{a}_{\phi}(\bm{z}_0) \cdot \nabla_{\bm{z}} \hat{Z}_{\lambda}(\bm{x}|\bm{z};\omega)\big|_{\bm{z}=\bm{z}_0}.
  \label{SMeq:multi_v_hat_definition}
\end{equation}
Differentiating \cref{SMeq:multi_Z_hat_equation} with respect to $\bm{z}$ along the direction $\bm{a}_{\phi}(\bm{z}_0)$ and setting $\bm{z}=\bm{z}_0$, we obtain
\begin{equation}
  (\omega - \mathcal{L}_{\lambda})\hat{v}_{\lambda,\phi}(\bm{x},\omega) = u_{\phi,\bm{z}_0}(\bm{x}),
  \label{SMeq:multi_v_hat_equation}
\end{equation}
where $u_{\phi,\bm{z}_0}(\bm{x})$ is the same local source introduced in \cref{SMeq:multi_vlambda_equation}.

As in the steady-state case, the derivative of the generator with respect to $\lambda$ is still given by the rank-one structure in \cref{SMeq:multi_generator_derivative}. Differentiating \cref{SMeq:multi_v_hat_equation} with respect to $\lambda$, we find
\begin{equation}
  -\frac{d\mathcal{L}_{\lambda}}{d\lambda}\hat{v}_{\lambda,\phi} + (\omega - \mathcal{L}_{\lambda})\frac{d\hat{v}_{\lambda,\phi}}{d\lambda} = 0.
  \label{SMeq:multi_v_hat_derivative_1}
\end{equation}
Using \cref{SMeq:multi_generator_derivative}, this becomes
\begin{equation}
  (\omega - \mathcal{L}_{\lambda})\frac{d\hat{v}_{\lambda,\phi}}{d\lambda} = u_{\phi,\bm{z}_0}\,\ell_{\phi,\bm{z}_0}[\hat{v}_{\lambda,\phi}].
  \label{SMeq:multi_v_hat_derivative_2}
\end{equation}
Comparing \cref{SMeq:multi_v_hat_derivative_2} with \cref{SMeq:multi_v_hat_equation}, and using the invertibility of $\omega-\mathcal{L}_{\lambda}$ for $\operatorname{Re}\omega > 0$, we obtain
\begin{equation}
  \frac{d\hat{v}_{\lambda,\phi}(\bm{x},\omega)}{d\lambda} = \ell_{\phi,\bm{z}_0}[\hat{v}_{\lambda,\phi}]\,\hat{v}_{\lambda,\phi}(\bm{x},\omega).
  \label{SMeq:multi_v_hat_scalar_equation}
\end{equation}
Thus, for any reference value $\lambda_{\ast}$, there exists a scalar function $\hat{M}(\lambda,\omega)$ such that
\begin{equation}
  \hat{v}_{\lambda,\phi}(\bm{x},\omega) = \hat{M}(\lambda,\omega)\,\hat{v}_{\lambda_{\ast},\phi}(\bm{x},\omega).
  \label{SMeq:multi_v_hat_proportionality}
\end{equation}

We next derive the $\lambda$ dependence of the Laplace-transformed density itself. Differentiating \cref{SMeq:multi_p_hat_resolvent_equation} with respect to $\lambda$ yields
\begin{equation}
  -\frac{d\mathcal{L}_{\lambda}}{d\lambda}\hat{p}_{\lambda} + (\omega - \mathcal{L}_{\lambda})\frac{d\hat{p}_{\lambda}}{d\lambda} = 0.
  \label{SMeq:multi_p_hat_derivative_1}
\end{equation}
Using \cref{SMeq:multi_generator_derivative}, we obtain
\begin{equation}
  (\omega - \mathcal{L}_{\lambda})\frac{d\hat{p}_{\lambda}}{d\lambda} = u_{\phi,\bm{z}_0}\,\ell_{\phi,\bm{z}_0}[\hat{p}_{\lambda}].
  \label{SMeq:multi_p_hat_derivative_2}
\end{equation}
Comparing \cref{SMeq:multi_p_hat_derivative_2} with \cref{SMeq:multi_v_hat_equation}, we find
\begin{equation}
  \frac{d\hat{p}_{\lambda}(\bm{x},\omega)}{d\lambda} = \ell_{\phi,\bm{z}_0}[\hat{p}_{\lambda}]\,\hat{v}_{\lambda,\phi}(\bm{x},\omega).
  \label{SMeq:multi_p_hat_response}
\end{equation}
Substituting \cref{SMeq:multi_v_hat_proportionality} into \cref{SMeq:multi_p_hat_response}, we conclude that the Laplace-transformed density also evolves along a one-dimensional affine manifold,
\begin{equation}
  \hat{p}_{\lambda}(\bm{x},\omega) = \hat{p}_{\lambda_{\ast}}(\bm{x},\omega) + \hat{s}(\lambda,\omega)\,\hat{v}_{\lambda_{\ast},\phi}(\bm{x},\omega),
  \label{SMeq:multi_p_hat_affine}
\end{equation}
where $\hat{s}(\lambda,\omega)$ is a scalar function determined by $d\hat{s}(\lambda,\omega)/d\lambda = \ell_{\phi,\bm{z}_0}[\hat{p}_{\lambda}]\,\hat{M}(\lambda,\omega)$.
Therefore, for any two observation points $\bm{x}_1$ and $\bm{x}_2$ such that $\hat{v}_{\lambda_{\ast},\phi}(\bm{x}_2,\omega)\neq 0$, one has
\begin{equation}
  \hat{p}_{\lambda}(\bm{x}_1,\omega) = \hat{\chi}_{\bm{x}_1\bm{x}_2}(\omega)\,\hat{p}_{\lambda}(\bm{x}_2,\omega) + \hat{\gamma}_{\bm{x}_1\bm{x}_2}(\omega),
  \label{SMeq:multi_laplace_density_mutual_linearity}
\end{equation}
where
\begin{equation}
  \hat{\chi}_{\bm{x}_1\bm{x}_2}(\omega) \equiv \frac{\hat{v}_{\lambda_{\ast},\phi}(\bm{x}_1,\omega)}{\hat{v}_{\lambda_{\ast},\phi}(\bm{x}_2,\omega)},
  \qquad
  \hat{\gamma}_{\bm{x}_1\bm{x}_2}(\omega) \equiv \hat{p}_{\lambda_{\ast}}(\bm{x}_1,\omega) - \hat{\chi}_{\bm{x}_1\bm{x}_2}(\omega)\hat{p}_{\lambda_{\ast}}(\bm{x}_2,\omega).
  \label{SMeq:multi_laplace_density_coefficients}
\end{equation}

We now turn to state-current observables. Define
\begin{equation}
  Q_{m,\lambda}(t) \equiv \int d\bm{x}\,a_m(\bm{x})p_{\lambda}(\bm{x},t) + \int d\bm{x}\,\bm{b}_m(\bm{x}) \cdot \bm{j}_{\lambda}(\bm{x},t),
  \qquad
  m = 1,2,
  \label{SMeq:multi_Qm_time_average}
\end{equation}
where
\begin{equation}
  \bm{j}_{\lambda}(\bm{x},t) \equiv \bm{\mathcal{J}}_{\lambda}p_{\lambda}(\bm{x},t).
  \label{SMeq:multi_current_time_dependent}
\end{equation}
We then define the Laplace transform of the time-dependent observable by
\begin{equation}
  \hat{Q}_m(\omega)\big|_{\lambda} \equiv \int_0^{\infty} e^{-\omega t} Q_{m,\lambda}(t)\,dt.
  \label{SMeq:multi_Qm_hat_definition}
\end{equation}
Assume again that the observables satisfy the multidimensional excluded-point condition
\begin{equation}
  \bm{b}_m(\bm{z}_0) \cdot \bm{a}_{\phi}(\bm{z}_0) = 0,
  \qquad
  m = 1,2.
  \label{SMeq:multi_Qm_hat_excluded_point}
\end{equation}
Then, using the same decomposition as in \cref{SMeq:multi_current_split}, the contact contribution vanishes exactly, and the same adjoint manipulation as in \cref{SMsec:multi_observable} gives
\begin{equation}
  \hat{Q}_m(\omega)\big|_{\lambda} = \int d\bm{x}\,c_m(\bm{x})\hat{p}_{\lambda}(\bm{x},\omega),
  \qquad
  c_m(\bm{x}) \equiv a_m(\bm{x}) + \bm{\mathcal{J}}_0^{\dagger}\bm{b}_m(\bm{x}).
  \label{SMeq:multi_Qm_hat_density_functional}
\end{equation}
Substituting \cref{SMeq:multi_p_hat_affine} into \cref{SMeq:multi_Qm_hat_density_functional}, we obtain
\begin{equation}
  \hat{Q}_m(\omega)\big|_{\lambda} = \hat{A}_m(\omega) + \hat{s}(\lambda,\omega)\hat{B}_m(\omega),
  \label{SMeq:multi_Qm_hat_affine}
\end{equation}
where
\begin{equation}
  \hat{A}_m(\omega) \equiv \int d\bm{x}\,c_m(\bm{x})\hat{p}_{\lambda_{\ast}}(\bm{x},\omega),
  \qquad
  \hat{B}_m(\omega) \equiv \int d\bm{x}\,c_m(\bm{x})\hat{v}_{\lambda_{\ast},\phi}(\bm{x},\omega).
  \label{SMeq:multi_Qm_hat_affine_coefficients}
\end{equation}
Therefore, for any two such observables with $\hat{B}_2(\omega)\neq 0$, eliminating the common scalar $\hat{s}(\lambda,\omega)$ gives
\begin{equation}
  \hat{Q}_1(\omega)\big|_{\lambda} = \hat{\chi}_{12}(\omega)\,\hat{Q}_2(\omega)\big|_{\lambda} + \hat{\gamma}_{12}(\omega),
  \label{SMeq:multi_laplace_observable_mutual_linearity}
\end{equation}
with
\begin{equation}
  \hat{\chi}_{12}(\omega) \equiv \frac{\hat{B}_1(\omega)}{\hat{B}_2(\omega)},
  \qquad
  \hat{\gamma}_{12}(\omega) \equiv \hat{A}_1(\omega) - \frac{\hat{B}_1(\omega)}{\hat{B}_2(\omega)}\hat{A}_2(\omega).
  \label{SMeq:multi_laplace_observable_coefficients}
\end{equation}
Finally, the steady-state result is recovered from the small-$\omega$ limit. Since
\begin{equation}
  \hat{p}_{\lambda}(\bm{x},\omega) = \frac{\pi_{\lambda}(\bm{x})}{\omega} + O(1)
  \qquad
  (\omega \to 0^{+}),
  \label{SMeq:multi_small_omega_limit}
\end{equation}
the affine relations in \cref{SMeq:multi_laplace_density_mutual_linearity,SMeq:multi_laplace_observable_mutual_linearity} reduce to their steady-state counterparts \cref{SMeq:multi_density_mutual_linearity,SMeq:multi_observable_mutual_linearity}.

\section{Finite-width Correction Effect}
\label{SMsec:finite-width}

In the main text, mutual linearity is derived for ideal point perturbations of the form $\phi_{\lambda}(x)=\phi_0(x)+\lambda\delta(x-z_0)$. Here, we analyze how the result changes when the point perturbation is replaced by a narrow Gaussian packet,
\begin{subequations}
\begin{align}
  \phi_{\lambda,h}(x) &= \phi_0(x)+\lambda G_h(x-z_0), \\
  G_h(x-z_0) &= \frac{1}{h\sqrt{2\pi}}\exp\!\left[-\frac{(x-z_0)^2}{2h^2}\right].
\end{align}
\label{SMeq:gaussian_perturbation}
\end{subequations}
The Gaussian $G_h$ is normalized and has variance $h^2$.

The key observation is that Gaussian averaging can be expanded in powers of the width. For any sufficiently smooth test function $f(z)$,
\begin{equation}
  \int \rmd z\,G_h(z-z_0)f(z)
  =
  f(z_0)+\frac{h^2}{2}f''(z_0)+O(h^4).
  \label{SMeq:gaussian_test_function_expansion}
\end{equation}
This follows directly from the Taylor expansion $f(z)=f(z_0)+(z-z_0)f'(z_0)+(z-z_0)^2f''(z_0)/2+\cdots$ and the Gaussian moments
\begin{equation}
  \int \rmd z\,G_h(z-z_0)(z-z_0)=0,
  \qquad
  \int \rmd z\,G_h(z-z_0)(z-z_0)^2=h^2.
  \label{SMeq:gaussian_moments}
\end{equation}
Equivalently, in the distributional sense,
\begin{equation}
  G_h(z-z_0)
  =
  \delta(z-z_0)+\frac{h^2}{2}\partial_z^2\delta(z-z_0)+O(h^4),
  \label{SMeq:gaussian_distribution_expansion}
\end{equation}
where \cref{SMeq:gaussian_distribution_expansion} means that both sides have the same action on any smooth test function up to order $O(h^4)$.

We now apply this expansion to the response of an observable $Q_{m, \text{ss}, \lambda}$. For the ideal point perturbation at $z$, define the local response density
\begin{equation}
  S_m(z,\lambda) \equiv \frac{\delta Q_{m, \text{ss}, \lambda}}{\delta \phi(z)}
  \label{SMeq:local_response_density_definition}
\end{equation}
evaluated along the ideal perturbation path $\phi_\lambda(x)=\phi_0(x)+\lambda\delta(x-z_0)$. For density values,
\begin{equation}
  S_m(z,\lambda)
  =
  N_{\phi,\lambda}(z)v_{\lambda,z}(x_m),
  \label{SMeq:density_local_response_density}
\end{equation}
where $v_{\lambda,z}(x)=\partial_z Z_{\lambda}(x|z)$. For an excluded-point state-current observable, the contact contribution is absent and
\begin{equation}
  S_m(z,\lambda)
  =
  N_{\phi,\lambda}(z)\int \rmd x\,c_{Q_m}(x)v_{\lambda,z}(x),
  \label{SMeq:observable_local_response_density}
\end{equation}
provided that the current weight does not probe the perturbed region.

Under the Gaussian perturbation, the derivative with respect to $\lambda$ is a Gaussian average of local responses,
\begin{equation}
  \frac{\rmd Q_{m, \text{ss}, \lambda}^{(h)}}{\rmd \lambda} = \int \rmd z\,G_h(z-z_0)S_m(z,\lambda).
  \label{SMeq:gaussian_response_average}
\end{equation}
This correction can be absorbed into the coefficient of the leading $O(h^2)$ term below. Keeping the leading explicit Gaussian-averaging contribution, \cref{SMeq:gaussian_test_function_expansion} gives
\begin{equation}
  \frac{\rmd Q_{m, \text{ss}, \lambda}^{(h)}}{\rmd \lambda} = S_m(z_0,\lambda) + \frac{h^2}{2}\partial_z^2S_m(z,\lambda)\big|_{z=z_0} + O(h^4).
  \label{SMeq:gaussian_response_expansion}
\end{equation}

For $h=0$, exact mutual linearity gives
\begin{equation}
  Q_{1, \text{ss}, \lambda}^{(0)}(\lambda) = \chi_{12} Q_{2, \text{ss}, \lambda}^{(0)}(\lambda) + \gamma_{12}.
  \label{SMeq:ideal_mutual_linearity_for_finite_width}
\end{equation}
Differentiating \cref{SMeq:ideal_mutual_linearity_for_finite_width} with respect to $\lambda$ yields
\begin{equation}
  S_1(z_0,\lambda) = \chi_{12} S_2(z_0,\lambda).
  \label{SMeq:ideal_response_proportionality_at_z0}
\end{equation}
Define the affine residual for the finite-width perturbation as
\begin{equation}
  Y_{12}^{(h)}(\lambda)
  \equiv
  Q_{1, \text{ss}, \lambda}^{(h)}(\lambda) - \chi_{12} Q_{2, \text{ss}, \lambda}^{(h)}(\lambda) - \gamma_{12}.
  \label{SMeq:finite_width_residual_definition}
\end{equation}
Using \cref{SMeq:gaussian_response_expansion,SMeq:ideal_response_proportionality_at_z0}, we find
\begin{equation}
  \frac{\rmd Y_{12}^{(h)}}{\rmd \lambda} = \frac{h^2}{2} \left[ \partial_z^2 S_1(z,\lambda) - \chi_{12} \partial_z^2 S_2(z,\lambda) \right]_{z=z_0} + O(h^4).
  \label{SMeq:finite_width_residual_derivative}
\end{equation}
Choosing the same reference point at $\lambda=0$ for the ideal and finite-width perturbations, $Y_{12}^{(h)}(0) = 0$, integration over $\lambda$ gives
\begin{equation}
  Y_{12}^{(h)}(\lambda) = h^2 \Delta_{12}(\lambda) + O(h^4),
  \label{SMeq:finite_width_residual_scaling}
\end{equation}
where
\begin{equation}
  \Delta_{12}(\lambda) = \frac12\int_0^\lambda \rmd \lambda'\, \left[ \partial_z^2 S_1(z,\lambda') - \chi_{12} \partial_z^2 S_2(z,\lambda') \right]_{z=z_0}.
  \label{SMeq:finite_width_delta_expression}
\end{equation}
Equivalently,
\begin{equation}
  Q_{1, \text{ss}, \lambda}^{(h)}(\lambda) = \chi_{12} Q_{2, \text{ss}, \lambda}^{(h)}(\lambda) + \gamma_{12} + h^2 \Delta_{12}(\lambda)+O(h^4).
  \label{SMeq:finite_width_affine_error}
\end{equation}

For current-like observables, the condition $b(z_0)=0$ for an ideal perturbation should be strengthened for a finite-width perturbation. The current weight $b(x)$ should vanish, or at least be negligible, throughout the spatial region where $G_h(x-z_0)$ has appreciable weight. Under this finite-width excluded-region condition, the contact contribution remains subleading, and the leading deviation from mutual linearity is still governed by the $O(h^2)$ mode-mixing correction in \cref{SMeq:finite_width_affine_error}.

Finally, suppose the quality of the affine relation is quantified by the coefficient of determination $R^2$ over a fixed $\lambda$ interval. If the ideal mutual-linearity signal has a finite variance over this interval, while the vertical residual scales as $O(h^2)$, then the residual sum of squares scales as $O(h^4)$. Therefore,
\begin{equation}
  1 - R^2 = O(h^4).
  \label{SMeq:finite_width_R2_scaling}
\end{equation}
Thus, the affine residual is linear in the Gaussian variance $h^2$, while $1-R^2$ is quadratic in the variance.

\section{Details of numerical simulations}
\label{SMsec: numerical}

We simulate the F$_1$-ATPase model introduced in the main text on the unwrapped angular coordinate $\theta_t$,
\begin{equation}
  \dot{\theta}_t
  =
  \mu\left[-\partial_\theta U(\theta)\big|_{\theta=\theta_t}+F_{\mathrm{drive}}+\lambda G_h(\theta_t-z_0)\right]
  +
  \sqrt{2\mu T}\,\xi_t,
  \label{SMeq: f1_model_simulation}
\end{equation}
with $U(\theta)=U_0\cos(3\theta)$. The force and all observables are evaluated periodically with period $2\pi$, while the coordinate itself is evolved on the unwrapped real line.

The local perturbation is implemented as an area-normalized periodic Gaussian packet,
\begin{equation}
  G_h(\theta-z_0)
  =
  \sum_{n=-\infty}^{\infty}
  \frac{1}{h\sqrt{2\pi}}
  \exp\!\left[
  -\frac{(\theta-z_0+2\pi n)^2}{2h^2}
  \right],
  \label{SMeq: area_normalized_periodic_gaussian}
\end{equation}
which satisfies
\begin{equation}
  \int_0^{2\pi} \rmd\theta\,G_h(\theta-z_0)=1.
  \label{SMeq: gaussian_normalization}
\end{equation}
Thus $\lambda$ controls the integrated strength of the local torque perturbation. In the limit $h\to0$, $G_h(\theta-z_0)$ converges to the periodic delta function, and \cref{SMeq: f1_model_simulation} approaches the ideal local perturbation considered in the theory. In the numerical implementation, this periodic Gaussian is evaluated using the shortest periodic distance $d_{2\pi}(\theta,z_0)$,
\begin{equation}
  G_h(\theta-z_0)
  \simeq
  \frac{1}{h\sqrt{2\pi}}
  \exp\!\left[
  -\frac{d_{2\pi}(\theta,z_0)^2}{2h^2}
  \right],
  \label{SMeq: area_normalized_gaussian_shortest_distance}
\end{equation}
which is equivalent to keeping the dominant image contribution and is accurate for the widths used here.

The parameters are chosen as follows. We set $T=298\,\mathrm{K}$, so that $k_{\mathrm B}T=4.11\,\mathrm{pN}\,\mathrm{nm}$, take the rotational mobility $\mu=0.91\,\mathrm{rad}/(\mathrm{s}\,\mathrm{pN}\,\mathrm{nm})$, the potential amplitude $U_0=10k_{\mathrm B}T$, and the background torque $F_{\mathrm{drive}}=120\,\mathrm{pN}\,\mathrm{nm}$. The perturbed barrier is placed at $z_0=0$. We compare three widths, $h=0.05\,\mathrm{rad}$, $0.5\,\mathrm{rad}$, and $1.0\,\mathrm{rad}$, corresponding respectively to narrow, intermediate, and broad perturbations. Since the Gaussian is area-normalized, changing $h$ changes the spatial width of the perturbation while keeping its integrated strength fixed.

\bibliography{supplement}